\documentclass[journal, 12pt, draftclsnofoot, onecolumn]{IEEEtran}
\IEEEoverridecommandlockouts
\def\BibTeX{{\rm B\kern-.05em{\sc i\kern-.025em b}\kern-.08em
    T\kern-.1667em\lower.7ex\hbox{E}\kern-.125emX}}

\usepackage{amsthm,amsmath,amssymb,bm}
\usepackage{url}
\usepackage{verbatim,balance}
\usepackage{xcolor}
\usepackage{subfigure}
\usepackage{graphics,graphicx}
\usepackage{acronym}
\usepackage{upgreek}
\usepackage{setspace, textcomp, gensymb}
\usepackage{enumitem}
\usepackage[normalem]{ulem}

\begin{document}

\title{RIS-Enabled Smart Wireless Environments: \\Deployment Scenarios, Network Architecture, Bandwidth and Area of Influence}
\author{George C. Alexandropoulos, Dinh-Thuy Phan-Huy,  Kostantinos D. Katsanos, Maurizio Crozzoli, Henk Wymeersch, Petar Popovski, Philippe Ratajczak, Yohann B\'{e}n\'{e}dic, Marie-Helene Hamon, Sebastien Herraiz Gonzalez, \\Placido Mursia, Marco Rossanese, Vincenzo Sciancalepore, Jean-Baptiste Gros, Sergio Terranova, Gabriele Gradoni, Paolo Di Lorenzo, Moustafa Rahal, Benoit Denis, Raffaele D'Errico, Antonio Clemente, and Emilio Calvanese Strinati

 \thanks{This work has been supported by the EU H2020 RISE-6G project under grant number 101017011. Part of this paper has been presented in the Joint European Conference on Networks and Communications \& 6G Summit, Grenoble, France, June 2022 \cite{EuCNC}.}
\thanks{G. C. Alexandropoulos and K. D. Katsanos are with the Department of Informatics and Telecommunications, National and Kapodistrian University of Athens, Greece (e-mails: \{alexandg, kkatsan\}@di.uoa.gr). D.-T. Phan-Huy, P. Ratajczak, Y. B\'{e}n\'{e}dic, M.-H. Hamon, and S. Herraiz Gonzalez are with the Orange Innovation, France (emails: firstname.lastname@orange.com). M. Crozzoli is with the TIM, Italy (email: maurizio.crozzoli@telecomitalia.it). H. Wymeersch is with the Department of Electrical Engineering, Chalmers University of Technology, Sweden (email: henkw@chalmers.se). P. Popovski is with the Connectivity Section of the Department of Electronic Systems, Aalborg University, Denmark (e-mail: petarp@es.aau.dk). P. Mursia, M. Rossanese, and V. Sciancalepore are with NEC Laboratories Europe, Germany (emails: firstname.lastname@neclab.eu). J.-B. Gros is with Greenerwave, France (email: jean-baptiste.gros@greenerwave.com). S. Terranova and G. Gradoni are with University of Nottingham, United Kingdom (emails: firstname.lastname@nottingham.ac.uk). P. Di Lorenzo is with both the Department of Information Engineering, Electronics, and Telecommunications, Sapienza University of Rome and CNIT, Italy (email: paolo.dilorenzo@uniroma1.it). M. Rahal, B. Denis, R. D'Errico, A. Clemente, and E. Calvanese Strinati are with CEA-Leti, Universit\'{e} Grenoble Alpes, France (emails: firstname.lastname@cea.fr).}
 }

\maketitle

\begin{abstract} 
Reconfigurable Intelligent Surfaces (RISs) constitute the key enabler for programmable electromagnetic propagation environments, and are lately being considered as a candidate physical-layer technology for the demanding connectivity, reliability, localization, and sustainability requirements of next generation wireless networks. In this paper, we first present the deployment scenarios for RIS-enabled smart wireless environments that have been recently designed within the ongoing European Union Horizon 2020 RISE-6G project, as well as a network architecture integrating RISs with existing standardized interfaces. We identify various RIS deployment strategies and sketch the core architectural requirements in terms of RIS control and signaling, depending on the RIS hardware architectures and respective capabilities. Furthermore, we introduce and discuss, with the aid of simulations and reflectarray measurements, two novel metrics that emerge in the context of RIS-empowered wireless systems: the RIS bandwidth and area of influence. Their extensive investigation corroborates the need for careful deployment and planning of the RIS technology in future networks. 
\end{abstract}


\section{Introduction} \label{sec:intro}
The potential of Reconfigurable Intelligent Surfaces (RISs) for programmable Electro-Magnetic (EM) wave propagation \cite{WavePropTCCN} is lately gathering extensive academic and industrial interest, as an enabler for smart radio environments in the era of $6$-th Generation (6G) wireless networks \cite{RISE6G_COMMAG_all}. The RIS technology commonly refers to artificial planar structures with quasi-passive electronic circuitry, excluding any Radio-Frequency (RF) power amplification. An RIS, as a network element, is envisioned to be jointly optimized with conventional wireless transceivers~\cite{huang2019reconfigurable_all,risTUTORIAL2020} in order to significantly boost wireless communications in terms of coverage, spectral and energy efficiencies, reliability, and security, while satisfying regulated EM Field (EMF) emissions.

The most advanced RIS prototypes to date are built as antenna arrays operating in passive mode and including tunable unit cells, and can be seen as either reflective or transmissive surfaces \cite{alexandg_2021_all}. Specifically, a reflective RIS operates as an EM mirror, where an incident wave is reflected towards the desired direction (typically anomalous, in the sense that this direction can diverge from the specular direction dictated by geometrical optics) with specific radiation and polarization characteristics. On the other hand, a transmissive RIS operates as a lens or a frequency selective surface, where the incident field is manipulated (by filtering, polarization change, beam splitting, etc.) and/or phase shifted, and re-radiated so as to control the refraction of impinging plane waves \cite{Esposti2022}. It is noted that, when the RIS unit elements have both size and spacing lower than $1/10$-th of the communication operating wavelength, RISs are also termed as metasurfaces \cite{Glybovski_2016_all}. Although RISs have great potential to implement advanced EM wave manipulations, mainly simple functionalities, such as electronic beam steering and multi-beam scattering, have been demonstrated in the literature. Very recently, there has been increasing interest in the design of RIS hardware architectures \cite{Tsinghua_RIS_Tutorial}, in response to the RIS potential for wireless networking and the resulting challenges. Those architectures endow surfaces with either minimal reception RF chains \cite{alexandropoulos2021hybrid_all,HRIS_all,Alexandropoulos_2020a} or minimal power amplifiers \cite{Amplifying_RIS_all}, enabling more efficient channel parameters' estimation or extension of the RIS coverage area, respectively. Those capabilities can actually enable \emph{sensing}, where RISs can act standalone or facilitate network-wide orchestration via minimal required control with the rest of the network \cite{alexandropoulos2021hybrid_all}, or establish relay-like \emph{amplification} of the reflected signals in order to mitigate severe multiplicative path-loss.

The RIS-enabled programmability of information-bearing wireless signal propagation has recently gave birth to the concept of the \textit{Wireless Environment as a Service} \cite{rise6g_all}, which was introduced within the European Union (EU) Horizon 2020 (H2020) RISE-6G project\footnote{See \url{https://RISE-6G.eu} for more information.}. This environment offers dynamic radio wave control for trading off high-throughput communications, Energy Efficiency (EE), localization accuracy, and secrecy guarantees over eavesdroppers, while accommodating specific regulations on spectrum usage and restrained EMF emissions. However, the scenarios under which RISs will offer substantial performance improvement over conventional network architectures have not been fully identified. In this article, capitalizing on the latest studies within the RISE-6G project, we try to fill this gap and present various deployment scenarios for RIS-enabled smart wireless environments, while discussing their core architectural requirements in terms of RIS control and signaling, depending on the RIS hardware architectures and their respective capabilities. We also introduce two key novel metrics that emerge in the context of  RIS-enabled smart wireless environments: the \textit{Bandwidth of Influence (BoI)} and the \textit{Area of Influence (AoI)} of RISs, which corroborate the need for careful deployment and planning of the technology.

The remainder of this article is organized as follows. In Sec.~\ref{sec:all_scenarios}, various RIS deployment scenarios are presented, which are grouped based on the application and respective performance objective, namely, connectivity and reliability, localization and sensing, as well as sustainability and secrecy.   
In Sec.~\ref{sec:RIS_deploym}, we discuss the integration of RISs in standardized wireless networks and present a network architecture for RIS-enabled smart wireless environments. In this section, we also introduce two novel challenges imposed by RISs, namely, their BoI and AoI. Sections~\ref{sec:BoI_Char} and~\ref{sec:AoI_char} respectively include the characterization of these two challenges, both through mathematical definitions and computer simulations as well as experimental investigations with RISE-6G's fabricated RISs. Finally, Sec.~\ref{sec:concl} concludes the paper.

\section{RIS Deployment Scenarios} \label{sec:all_scenarios}
In this section, we present general scenarios for connectivity and reliability with RISs, localization and sensing applications with RISs, as well as RIS deployments scenarios for sustainability and secrecy. In most of the scenarios, we will distinguish between the consideration of RISs for boosting the performance of established wireless systems and for enabling wireless applications, which are otherwise infeasible. 

\subsection{Connectivity and Reliability} \label{sec:conn_rel}
Conventional network scenarios impose communication performance to be achieved via uncontrolled wireless propagation media, where the network is required to be tuned to one of the available service modes in a time orthogonal manner, offering non-focused areas of harmonized and balanced performance. However, this might result in resource inefficiency and huge complexity. Conversely, the envisioned smart wireless environment, built upon RISs, will enable the granting of highly localized quality of experience and specific service types to the end users \cite{rise6g_all}. In fact, such pioneering network paradigm aims at going one step beyond the classical $5$-th Generation (5G) use cases, by provisioning performance-boosted areas as dynamically designed regions that can be highly localized, offering customized high-resolution manipulation of radio wave propagation to meet selected performance indicators. Although we do not cover the details in this arcticle, it is important to note that one of the central architectural elements of RIS-empowered wireless systems is the control channel \cite{RISE6G_COMMAG_all}. The rate and latency of this channel are crucial for the timely and efficient configuration of RIS-enabled smart wireless environments, and should be in accord with the overall system requirements. 

For conventional system settings and strategies, based on a qualitative analysis of both localization feasibility (including possibly high-level identifiability considerations) and expected performance, a system engineer needs to determine where and how RISs can improve connectivity. In all scenarios described in the sequel, we consider that RISs are in reflecting operating mode. It is noted, however, that a recent line of research \cite{alexandropoulos2021hybrid_all,HRIS_all,Alexandropoulos_2020a} focuses on RISs that can also receive signals to, for instance, perform in-band channel estimation or localization. This capability is expected to have an impact on the control channels for orchestrating smart wireless environments. Nevertheless, the deployments scenarios of sensing RIS will be similar to those of almost passive RISs, as long as the adoption of the former targets wave propagation control via RF sensing. Focusing on the downlink case, we next provide a taxonomy of  RIS-empowered connectivity and reliability scenarios; the uplink case can be readily extended following analogous approach.
\begin{figure}[!t]
    \centering
    \subfigure[Connectivity and reliability boosted by a single RIS. ]{\includegraphics[scale=0.55]{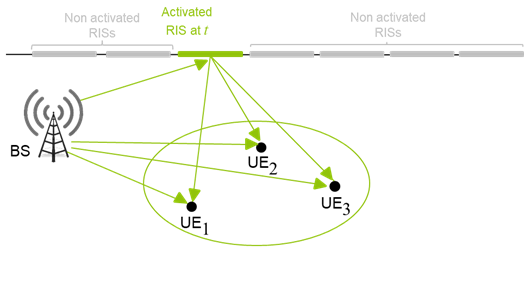}} \\
    \subfigure[RIS-empowered downlink communications of two BS-UE pairs, where each RIS can be controlled individually by each pair.]{\includegraphics[scale=0.85]{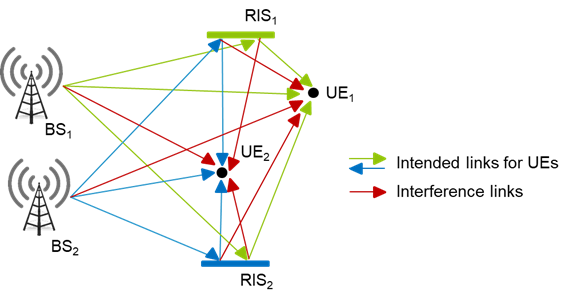}}
    \caption{Illustration of the connectivity and reliability scenarios~\ref{sec:single_RIS} and~\ref{sec:multi_RIS}.}
    \label{fig:scenarios_1_2}
\end{figure}

\subsubsection{Connectivity and Reliability Boosted by a Single RIS} \label{sec:single_RIS}
In this scenario, there exist(s) direct communication link(s) between the multi-antenna Base Station (BS) and the single- or multi-antenna User Equipments (UEs). However, connectivity can be further boosted efficiently via a single RIS, as shown in Fig.~\ref{fig:scenarios_1_2}a. The phase configuration/profile of the RIS can be optimized via a dedicated controller, who interacts with the BS, for the desired connectivity and reliability levels.

\subsubsection{Connectivity and Reliability Boosted by Multiple Individually Controlled RISs} \label{sec:multi_RIS}
In this scenario, multiple BSs aim at boosting connectivity and reliability with their respective UEs. The deployed RISs are assigned to BS-UE pairs, and each pair is capable of controling and optimizing the phase profile of its assigned individual RIS(s), as illustrated in Fig.~\ref{fig:scenarios_1_2}b. Since the action of an RIS can affect the propagation for links that are not intended to be controlled by that RIS, the result is that RIS operation interferes with the link operation. Unlike the usual wireless interference, which is additive, this can be seen as a \emph{multiplicative interference} that needs to be handled carefully. Examples of interference management via a Software Defined Radio (SDR) setup with two RISs, targeting channel hardening of the received signals at UE terminals have been lately presented~\cite{lodro2021reconfigurable,gros2022multi,lodro2022experimental}. The characterization of the capacity limits in multi-RIS-empowered wireless communications, and the practical schemes to achieve it, constitutes also an active area of research \cite{DRL_automonous_RIS, RIS_placement}.

\subsubsection{Connection Reliability Enabled by Multiple RISs} \label{sec:multi_RIS2}
In this scenario, the direct link(s) between the multi-antenna BS and the single- or multi-antenna UE(s) is (are) blocked, and connectivity is feasibile only via a single or multiple RISs, as shown in Fig.~\ref{fig:scenarios_3_4}a. Similar to scenario 2.1, the phase profile(s) of the RIS(s) can be optimized for desired reliability levels, requiring, however, forms of coordination among the different BSs (e.g., possibly similar to the coordinated beamforming mechanism in LTE-Advanced~\cite{ltea_all}).

\subsubsection{Connectivity and Reliability Boosted by a Single Multi-Tenant RIS} \label{sec:multi_ten_RIS}
\begin{figure}[!t]
    \centering
    \subfigure[RIS-aided systems where connectivity is enabled by multiple RISs.]{\includegraphics[trim={0cm 0.45cm 0cm 0cm},clip,scale=0.85]{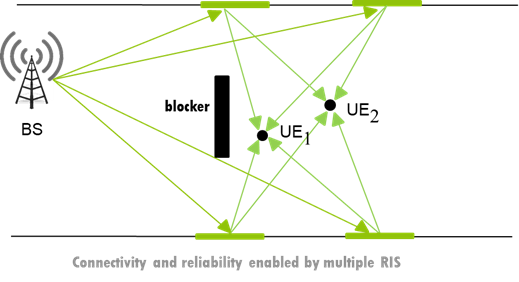}} \\
    \subfigure[A multi-tenancy scenario with two BS-UE pairs and a shared RIS that is optimized to simultaneously boost reliable communications.]{\includegraphics[scale=0.85]{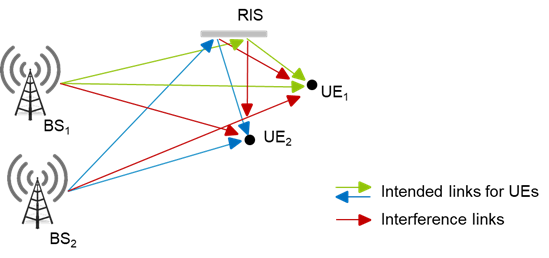}}
    \caption{Illustration of the connectivity and reliability scenarios~\ref{sec:multi_RIS2} and~\ref{sec:multi_ten_RIS}.}
    \label{fig:scenarios_3_4}
\end{figure}

For this scenario, we consider pairs of BS-UE(s) and a single RIS, as depicted in Fig.~\ref{fig:scenarios_3_4}b. The RIS is now considered as a shared resource, dynamically controlled by the infrastructure and commonly accessed by the BS-UE(s) pairs. The phase profile of the RIS can be commonly optimized by the BSs to serve their own UE(s) simultaneously. Alternatively, the control of the RIS may be time-shared among the BS-UE(s) pairs. Of course, the control channel envisioned by this scenario will have to be thoroughly investigated in future activities. A special case of this scenario is the one which considers a setup where the communication is enabled by multiple cellular BSs, each one serving a distinct set of UEs. When the UE(s) move across the cell boundaries of two or more BSs, they might change their serving BS(s) frequently resulting in frequent handover. Shared RISs among BSs can be placed in the cell boundaries in order to dynamically extend the coverage of the serving BSs, thereby reducing the number of handovers.

\subsubsection{RIS-Empowered Multi-Access Edge Computing} \label{sec:mec_RIS}
\begin{figure*}[!t]
\centering
  \includegraphics[scale=0.55]
  {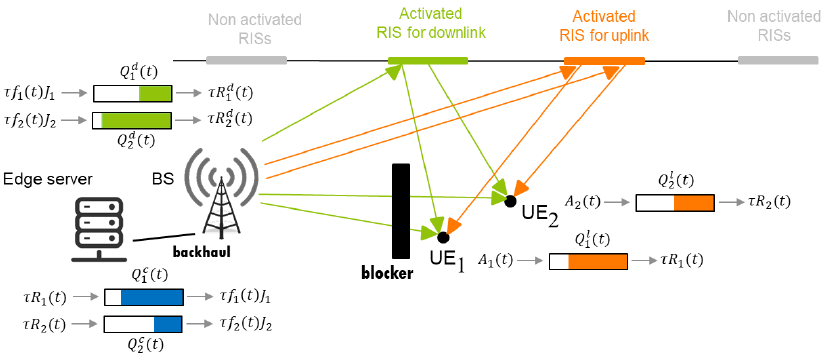}
  \caption{An RIS-enabled multi-access edge computing system \cite{Lyapunov_RIS_MEC_2023}. At each time slot $t$, new offloading requests are generated by the UEs, which are handled by a dynamic queueing mechanism that accounts for communication and processing delays, and targets at jointly optimizing communication and computation resources, while guaranteeing low-latency and high-accuracy requirements. $\tau$ denotes the time slot duration and $f_k(t)$ is the CPU frequency allocated by the edge server to UE$_k$ with $k \in \{1,2\}$. $R_k(t)$ and $R_k^d(t)$ stand for the uplink and downlink rates, respectively, while parameter $J_k$ depends on the application offloaded by device $k$. $Q_k^l$ denotes the local queue update, $Q_k^c$ is the computation queue at the edge server, and $Q_k^d$ is the downlink communication queue.}
  \label{fig:scenario_MEC}
\end{figure*}

With the advent of beyond 5G networks, mobile communication systems are evolving from a pure communication framework to enablers of a plethora of new services, such as Internet of Things (IoT) and autonomous driving. These new services present very diverse requirements, and they generally involve massive data processing within low end-to-end delays. In this context, a key enabler is Multi-Access Edge Computing (MEC), whose aim is to move cloud functionalities (e.g., computing and storage resources) at the edge of the wireless network, to avoid the relatively long and highly variable delays necessary to reach centralized clouds. MEC-enabled networks allow UEs to offload computational tasks to nearby processing units or edge servers, typically placed close to the access points, to run the computation on the UEs’ behalf. However, moving towards millimeter-wave and THz communications, poor channel conditions due to mobility, dynamics of the environment, and blocking events, might severely hinder the performance of MEC systems. In this context, a strong performance boost can be achieved with the advent of RISs, which enable programmability and adaptivity of the wireless propagation environment, by dynamically creating service boosted areas where EE, latency, and reliability can be traded to meet temporary and location-dependent requirements of MEC systems  \cite{bai2021reconfigurable_all,battiloro2022lyapunov_all,Lyapunov_RIS_MEC_2023}. Figure~\ref{fig:scenario_MEC} depicts an RIS-enabled MEC system, where RISs are deployed to establish the wireless connections of two UEs with the edge server, via a BS which is backhaul connected with it. 

\subsection{Localization and Sensing} \label{sec:local_sens}
The RIS technology is expected to enable advanced sensing and localization techniques for environment mapping, motion detection, opportunistic channel sounding, and passive radar capabilities applied to industrial (e.g., smart factory), high user-density (e.g. train stations), and indoor (e.g. augmented/mixed reality) environments. 
The expected benefits can take various forms, depending on the RIS operating mode (i.e., reflect, refract, transmit, relay, etc. \cite{Tsinghua_RIS_Tutorial}), and they can be classified in terms of \textit{i}) {enabled localization} (i.e., making localization feasible where any conventional system  \cite{delPeral_Rosado_2018_all,Keating_2019_all} fails);
 \textit{ii}) {boosted localization} (i.e., improving the localization performance); and \textit{iii})
{low-profile localization} (i.e., localization 
requiring much lower resources in comparison with the conventional system). 

\begin{figure}[!t]
    \centering
    \includegraphics[width=0.9\textwidth]{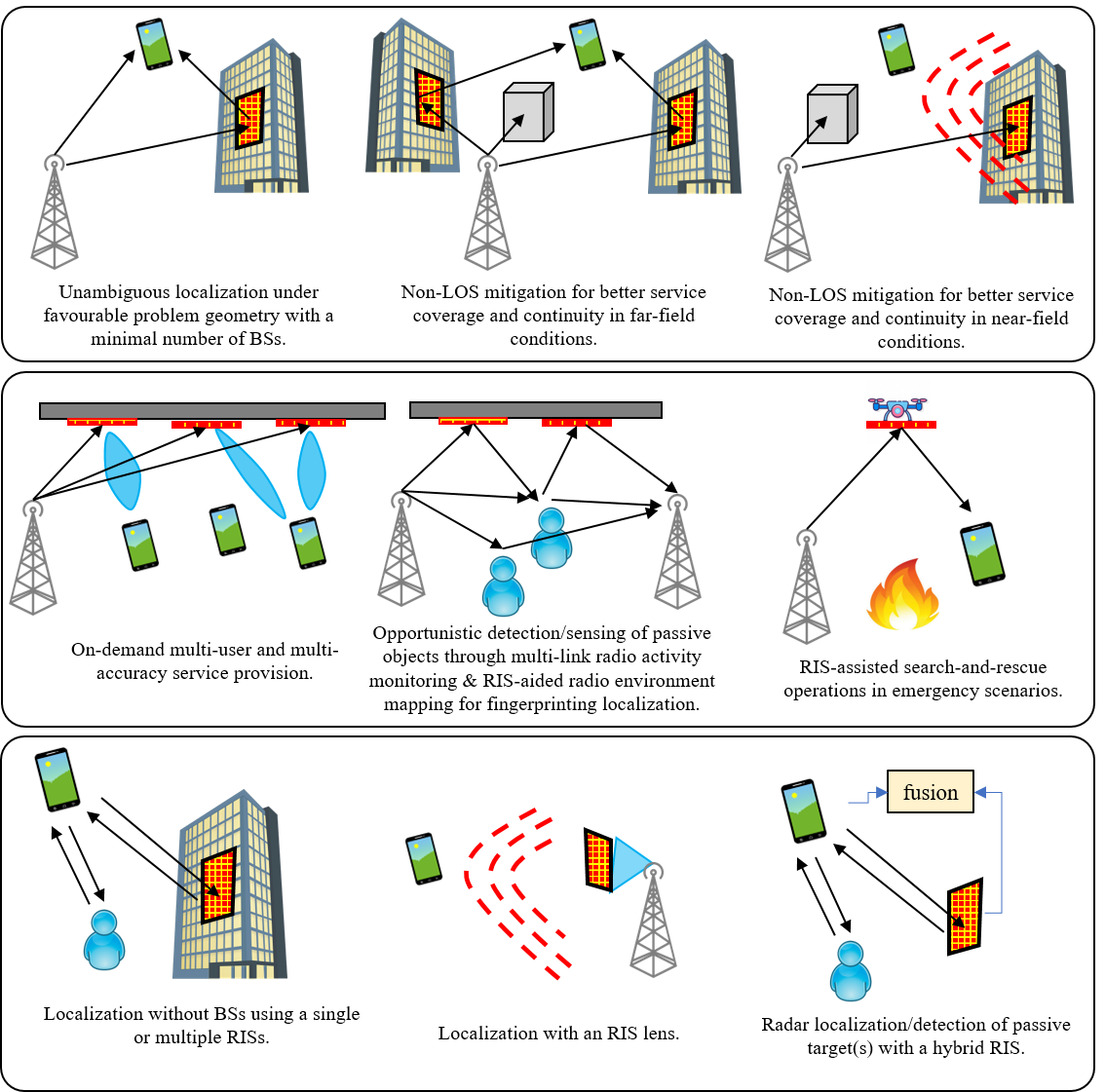}
    \caption{Scenarios of RIS-aided and RIS-enabled localization. }
    \label{fig:RISlocalization}
\end{figure}

In the following, ten generic scenarios where RISs provide performance benefits for localization are presented 
(refer to the relevant open technical literature for a more comprehensive overview \cite{Keykhosravi_2021_all, Abu_Shaban_2021_all, Keykhosravi_2021a_all,Keykhosravi_2023_VTM,Hui_smart_cities}). These scenarios are visualized in Fig.~\ref{fig:RISlocalization}.

\begin{itemize}
\item \emph{Unambiguous localization under favourable problem geometry with a minimal number of BSs:} \label{sec:unamb_minimal}
A single RIS enables localization \cite{Keykhosravi_2021_all}, while multi-RIS deployment settings \cite{Loc_Multiple_Passive_RISs_2023} can also contribute to boost performance, by a combination of time and angle measurements. 
\item \emph{Non-Line-of-Sight (NLoS) mitigation for better service coverage and continuity in far-field conditions:} \label{sec:nlos_far} Whenever the minimum number of BSs (anchors) in visibility is not fulfilled, the UE location can be estimated via narrowband signals received from two RISs. 
\item \emph{NLoS mitigation for better service coverage and continuity in near-field conditions:} \label{sec:nlos_near}
Extending the previous scenario, the UE location can be estimated via the signal received from one RIS even without any BS in visibility, when the user is in the near-field of the RIS. This allows to exploit signal wavefront curvature for direct positioning \cite{Abu_Shaban_2021_all,Rahal_2021_all}. 
\item \emph{On-demand multi-user and multi-accuracy service provision:} \label{sec:on_demand}
The deployment (and the selective control) of multiple RISs makes possible \textit{i}) the (on-demand) provision of various classes of localization services to different UEs sharing the same physical environment, depending on the needs they express locally/temporarily, while \textit{ii}) spatially controling both the localization accuracy and the geometric dilution of precision (GDoP) in different dimensions (i.e., both the sizes and orientations of the location uncertainty ellipses) \cite{Wymeersch_2020}. 
\item \emph{Opportunistic detection/sensing of passive objects through multi-link radio activity monitoring:} \label{sec:opport_detect}
Similar to standard range-Doppler analysis \cite{Li_2019_all}, this is possible by monitoring the time evolution of multipath profiles over a communication link between the BS and one or several UEs, harnessing the dynamic and selective control of RISs \cite{Lu_2021_all,Jiang_2021b_all,Buzzi_2021b_all, RIS_ISAC_2022}. In highly reverberant environments, multiple RISs can create configurational diversity through wavefront shaping, even with single-antenna single-frequency measurements \cite{alexandg_2021_all}.
\item \emph{RIS-assisted search-and-rescue operations in emergency scenarios:} \label{sec:emergency}
RISs may support and overcome the shadowing effect induced by rubble in such scenarios by building ad-hoc controllable propagation conditions for the cellular signals employed in the measurement process \cite{Albanese_2021_all}. 
In addition, lightweight and low-complexity RISs 
mounted on drones can be used to bring connectivity capabilities to hard-to-reach locations, supporting first responders \cite{Mursia_2021_all}. 
\item \emph{Localization without BSs using a single or multiple RISs:} \label{sec:noBSs}
In this scenario, wideband transmissions to a single passive RIS \cite{zerobs_all,Hyowon_AP_free} or angle estimations obtained at multiple RISs \cite{locrxris_all,Loc_R-RISs_2022}, having the architecture of \cite{Alexandropoulos_2020a}, can be combined to produce the estimation of UE(s) location(s).
\item \emph{RIS-aided radio environment mapping for fingerprinting localization:} \label{sec:fingerprinting}
RISs equipped with minimal reception circuitry for sensing \cite{Alexandropoulos_2020a} enable the cartography of the electromagnetic power spatial density in a specific area of interest, accentuating the location-dependent features of the RIS-enhanced radio signatures stored in the database. 
\item \emph{Localization with a RIS lens:} \label{sec:ris_lens}
In this scenario, the RIS is placed in front of a single-antenna transmitter. The user position is estimated at the UE side via the narrow-band received signal, exploiting wavefront curvature \cite{Abu_Shaban_2021_all}. 
\item \emph{Indoor and outdoor localization with simultaneously transmitting and reflecting RISs:} 
With the aid of a simultaneously transmitting and reflecting RISs~\cite{STAR_RIS}, an outdoor BS is capable of localizing an indoor user, in addition to an outdoor
user via uplink sounding reference signals~\cite{STAR-RIS_UAV_2022}.
\item \emph{Radar localization/detection of passive target(s) with simultaneously sensing and reflecting RISs:} \label{sec:radar}
Using the architecture of \cite{alexandropoulos2021hybrid_all,HRIS_all}, a radar is assisted by multiple hybrid RISs in order to localize/detect both static or moving target(s) \cite{Ghazalian_HRIS}, extending scenario 5. Hybrid RISs that can simultaneously sense (via dedicated reception RF chains or other signal sensing elements) and reflect providing the system with the capability to localize UEs as well as the radar itself. They can also be deployed for joint channel and direction estimation for ground-to-drone wireless communications~\cite{HRIS_UAV_2023}.
\end{itemize}

\subsection{Sustainability and Secrecy} \label{sec:sust_secur}
RIS-empowered smart radio environments are expected to improve the sustainability and security of 6G networks by focusing the energy towards the target UE, thus, reducing the amount of unnecessary radiations in non-intended directions in general, or in the directions of non-intended UEs (e.g., exposed UEs or eavesdroppers). More precisely, the following metrics can be improved: \textit{i}) the EE \cite{huang2019reconfigurable_all} (for instance, this metric can be defined as the attained data rate at the target UE divided by the BS transmit power); \textit{ii-a}) the self EMF Exposure Utility (S-EMFEU) (for instance, in uplink, this metric can be defined as the attained data rate of the target UE divided by its exposure to its own UE emissions, \textit{ii-b}) the inter EMFEU (I-EMFEU)  (for instance, this metric can be defined as the attained data rate at the target UE divided by the largest EMFE at another UE); and \textit{iii}) the Secrecy Spectral Efficiency (SSE) \cite{PLS_Kostas_all,PLS2022_counteracting} (for instance, this metric can be defined as the attained data rate at the target UE minus the data rate that is intercepted by the eavesdropper). A first already identified use case for such improvements \cite{rise6g_all} is a train station, where an orchestrated network of RISs can be optimally configured to maximize the aforementioned metrics, in small areas close to the surfaces. These areas can be advertized to customers as ``areas with high EE,'' ``areas with low EMF exposure,'' or ``areas with increased secrecy.''

\begin{figure*}[!t]
\centering
  \includegraphics[scale=0.255]
  {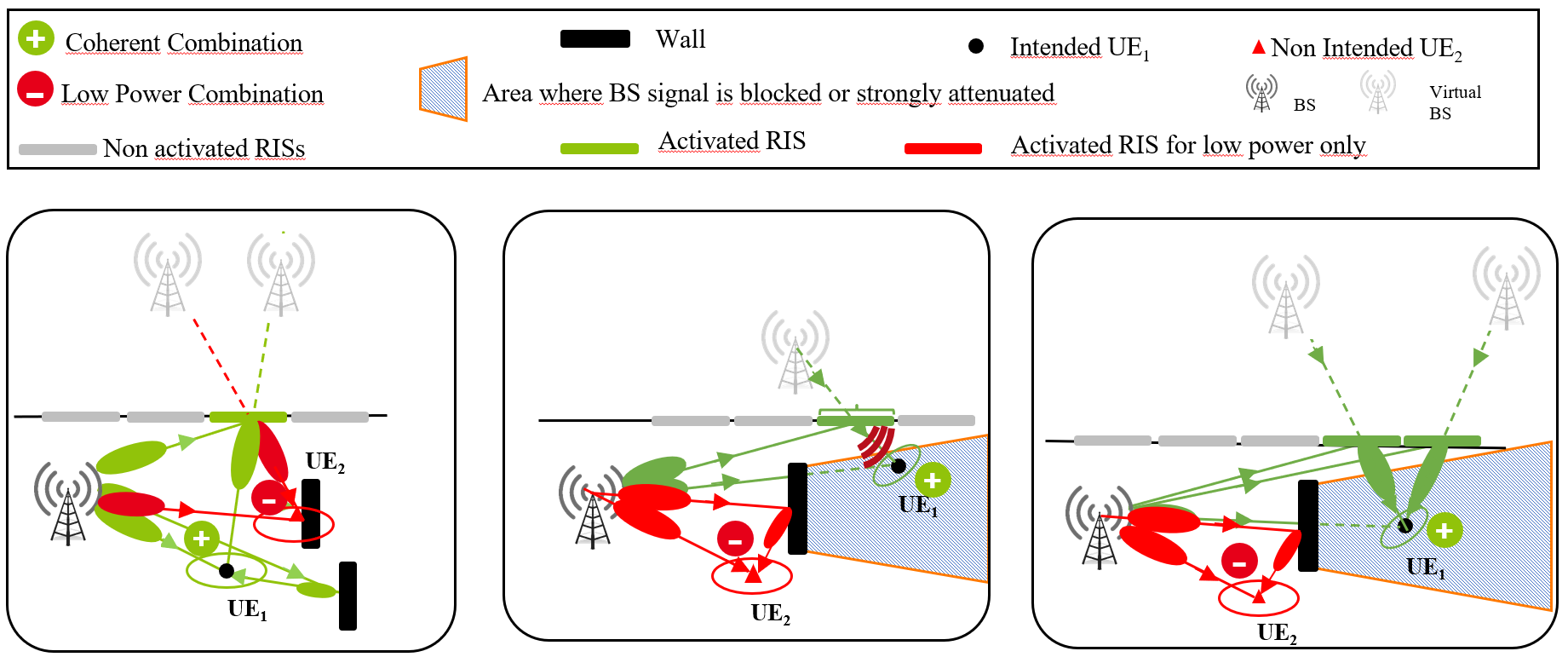}
  \caption{The three core scenarios for RIS-empowered sustainability and secrecy discussed in Section~\ref{sec:sust_secur}. In all scenarios, thanks to the RIS(s), UE$_2$ eavesdrops less (or is less exposed by) UE$_1$ downlink data because UE$_2$ is now on a location where propagation has been weakened artificially.}
  \label{fig:scenarios_4_3}
\end{figure*}

In contrast to conventional scenarios without deploying RISs, we illustrate in Fig.~\ref{fig:scenarios_4_3} examples of single-BS scenarios with RIS(s), where downlink transmit Beam Forming (BF) is used to boost the received power at the target intended UE and to reduce the received power at the non-intended UE; this can be achieved by exploiting the artificial shaping of the propagation channels thanks to RIS(s). When the non-intended UE is an exposed UE, the obtained link is a low EMF link, whereas when the non-intended UE is an eavesdropper, the obtained link is a secured link. In the figure, the advantages brought by an RIS(s) to reduce the received power at the non intended UE (whether it is an eavesdropper or an exposed UE), compared to the received power at the target UE, are illustrated for three types of propagation scenarios.
\begin{enumerate}
    \item \textit{The BS-to-intended-UE link is in LOS}: In this case (left figure), the RIS artificially adds propagation paths to the channel, which coherently combine with other ``natural'' paths to boost the received power at the target UE, and non-coherently combine with other ``physical'' paths to reduce the received power at the non-intended UE.
    \item \textit{The BS-to-intended-UE link is blocked by an obstacle and the intended UE is in the near-field of an RIS}: In this case (middle figure), the RIS artificially adds a propagation path to the existing ``physical'' paths to reduce the received power at the non-intended UE;
    \item \textit{The BS-to-intended-UE link is blocked by an obstacle and the intended UE is in the far-field of an RIS}: This case (right figure) is similar to the previous case except that several RISs may become useful to reduce the energy at the non-intended UE.
\end{enumerate}
In the absence of a non-intended UE, it is noted that the proposed system in Fig.~\ref{fig:scenarios_4_3} only boosts the received power at the target UE, and therefore, boosts the EE metric. Note also that similar scenarios for uplink can be derived with receive BF instead of transmit BF. Regarding EE, the main difference will be that the EE will be improved at the UE side. Regarding the EMF utility, the main difference will be that the target UE and the exposed UE will be the same entity, instead of being distinct entities. In this case, the RISs will help reducing the self-exposure of a UE transmitting data in the uplink with its own device. Furthermore, the previously described scenarios can be generalized to multi-BS scenarios, where several synchronized and coordinated BSs perform joint BF~\cite{ltea_all}.

\section{RIS Integration in Standardized Networks}\label{sec:RIS_deploym}
The deployment of wireless networks takes under consideration the following two basic principles of radio transmitters: a transmitter emits information-bearing signals in a limited spectrum bandwidth over a limited coverage area, depending on its hardware/software capabilities and various operation regulation factors. In addition, a wireless network operator ensures that the interference among its own BSs is minimal, by carefully planning their deployment and orchestrating their operation. More specifically, operators spatially distribute their BSs and access points to avoid overlaps among their limited coverage areas. Spectrum regulation ensures that interference among different network operators (with possibly overlapping coverage) is avoided, by allocating separate and limited spectrum bands, as well as defining spectrum masks and maximum out-of-band emission thresholds.

Very recently, RISs have been introduced as a new type of network node. A reflective RIS does not emit any signal, instead, it re-radiates the EM waves impinging on its panel~\cite{Marco2019}. As a result, such an RIS has no explicit coverage area nor explicit transmit bandwidth. In fact, the former will be determined by both the RIS capabilities and the sources whose radiated waves impinge on the RIS. For the latter, the RIS will have a frequency band where it can impact the reflections of its impinging signals, which will usually span a certain range of frequencies. Therefore, RISs cannot be trivially integrated neither in network planning nor spectrum regulation as usual BSs or other network devices. Interestingly, an RIS can influence the existing coverage of one or several networks. However, the influence of an RIS on an existing coverage is limited to its AoI. The influence of an RIS on its impinging signals is also limited in the frequency domain, as indicated by its BoI. In this section, we elaborate upon the two new concepts of RIS's AoI and BoI as well as their key roles for the deployment of the RIS technology in future wireless networks.
\begin{figure}[t!]
    \centering
    \includegraphics[scale=0.33]{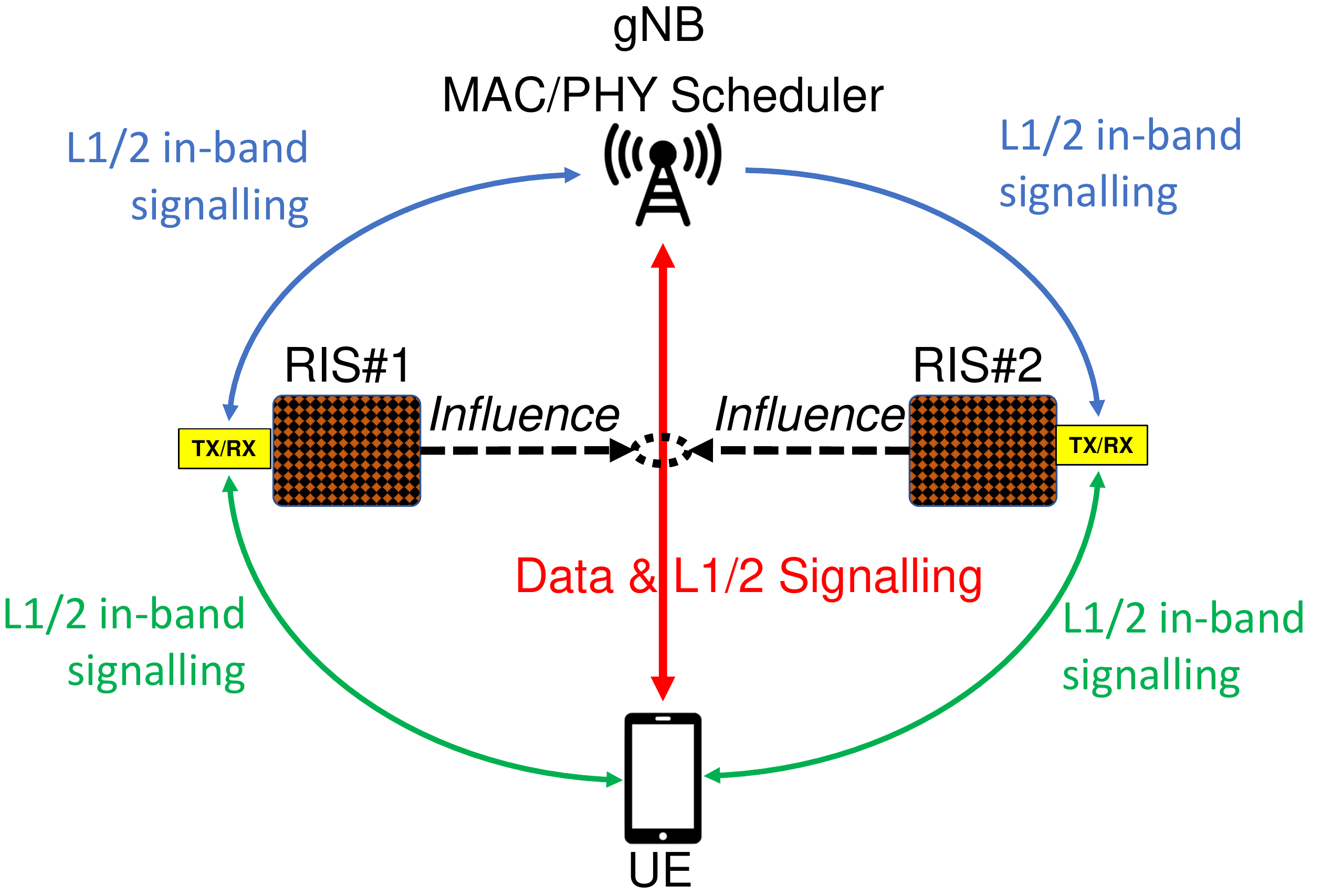}
    \caption{\small{An example integration scheme of two RISs in a 3GPP network, including a general interface among the different Radio Access Network (RAN) entities. Each RIS panel is assumed to be managed by a dedicated controller that includes a Transmitter (TX) and a Receiver (RX) modules for exchanging Layer 1 (L1) and Layer 2 (L2) in-band signaling.}}
    \label{fig:basicArch}
\end{figure}

\subsection{Network Architecture} \label{sec:arch}
In Fig.~\ref{fig:basicArch}, we present a simple integration scheme for RISs in 3GPP networks. As illustrated, the gNB communicates with a UE via the assistance of two reflective RISs, which cannot receive or emit any data (in the sense that they do no have such capability hardware-wise); extensions to RISs including active elements for reception and/or transmission are left for future investigations~\cite{Tsinghua_RIS_Tutorial}. Each RIS is assumed to only influence the quality of the data link between the UE and gNB, acting as a cost-efficient passive relay between those nodes. An RIS node comprises, apart from the reconfigurable reflection panel, a small and low-power signaling module, compactly known as the RIS controller, that is equipped with circuitry for managing the RIS phase configuration as well as transceiver radios to communicate with the gNB and UE, thus, enabling the RIS communication and interface with the rest of the network. In more detail, the RIS controller is responsible for generating the logical commands associated to the switching operations between the configurations/states of the RIS elements (e.g., predefined phase shifts), and can have different levels of complexity and capabilities, possibly embedding third-party applications to implement smart algorithms (e.g., the direction estimation of the impinging signals~\cite{locrxris_all}). The RIS controller may either accept requests from other elements in the network; in this case, it will simply act as an interface that configures the RIS elements based on external explicit instructions (this is an externally controlled RIS), or it may operate on its own (this is an autonomous RIS~\cite{alexandropoulos2021hybrid_all} that can perform self-optimization according to its own intention or as a response to a reflection-optimization request by an external cooperating node; the request can be realized with minimal interaction between that node and the RIS controller). The expected action time granularity for the operations of an RIS controller is envisioned to be between $20$ and $100$ milliseconds.

As depicted in Fig.~\ref{fig:basicArch}, the gNB and UE communicate with each RIS over-the-air through new radio interfaces and radio channels to be specified and embedded in the future New Radio (NR) standards. To this end, one can identify up to $8$ different types of L1/2 in-band signaling in this example integration figure:
\begin{itemize}
\renewcommand\labelitemi{--}
    \item C1: gNB-to-RIS control messages;
    \item C2: UE-to-RIS control messages;
    \item C3: RIS-to-gNB control messages;
    \item C4: RIS-to-UE control messages;
    \item P1: gNB-to-RIS pilot signals;
    \item P2: RIS-to-gNB pilot signals;
    \item P3: UE-to-RIS pilot signals; and
    \item P4: RIS-to-UE pilot signals.
\end{itemize}
It is noted that, with the proposed integration scheme and novel radio interfaces, many different types of basic RIS control can be implemented. In the following, we describe a representative example aiming the joint optimization of the BS transmit BF and the RIS reflective BF (i.e., RIS phase configuration/profile).

Assume the Time Division Duplex (TDD) mode and uplink-downlink channel reciprocity. In addition, the scheduler of the PHYsical (PHY) and Medium Access Control (MAC) layers considers the gNB as the master node for the two reflective RIS nodes, using C1 and P1 L1/2 signaling to control them. The steps for the intended joint BS/RIS BF optimization can be summarized as follows:
\begin{itemize}
\renewcommand\labelitemi{--}
    \item Step 1: The gNB sends requests to the RISs through C1 to switch between predefined reflective beams during a predefined sounding phase. 
    \item Step 2: During the sounding phase, the two RISs switch in an ordered and synchronized manner (thanks to synchronization signals over P1); the UE sends uplink pilots; and the gNB sounds the uplink channel for each reflective beam realized by the RISs. 
    \item Step 3: The gNB determines the jointly optimized transmit BF weights as well as the reflective BF configuration from the RISs. 
    \item Step 4: The gNB notifies through C1 the RISs to configure themselves according to the decided reflective BF configurations in the previous step. 
    \item Step 5: During the data transmission phase, the gNB transmits data using the best transmit BF weights. This data is received by the UE through a two-RISs-parameterized signal propagation channel that is subject to the desired influence of the two involved RISs.
\end{itemize}
Similarly, to allow control of any of the RISs by the UE, at least C2 and P2 signaling are necessary. C3 and C4 control signals could be also added to enable the RIS (actually, its controller) to answer requests from the UE and/or the gNB, and send acknowledgment or negative acknowledgment. The pilots signals P1 to P4 can be sent/received from/by each RIS controller. Hence, they cannot be used to sound the channel between the unit cells of each RIS and the gNB or UE. However, those pilots are necessary to estimate the propagation channel used by the L1 control signaling, facilitating its equalization, and consequently, signal demodulation.
\begin{figure*}[t!]
    \centering
    \includegraphics[clip,width=0.99\textwidth]{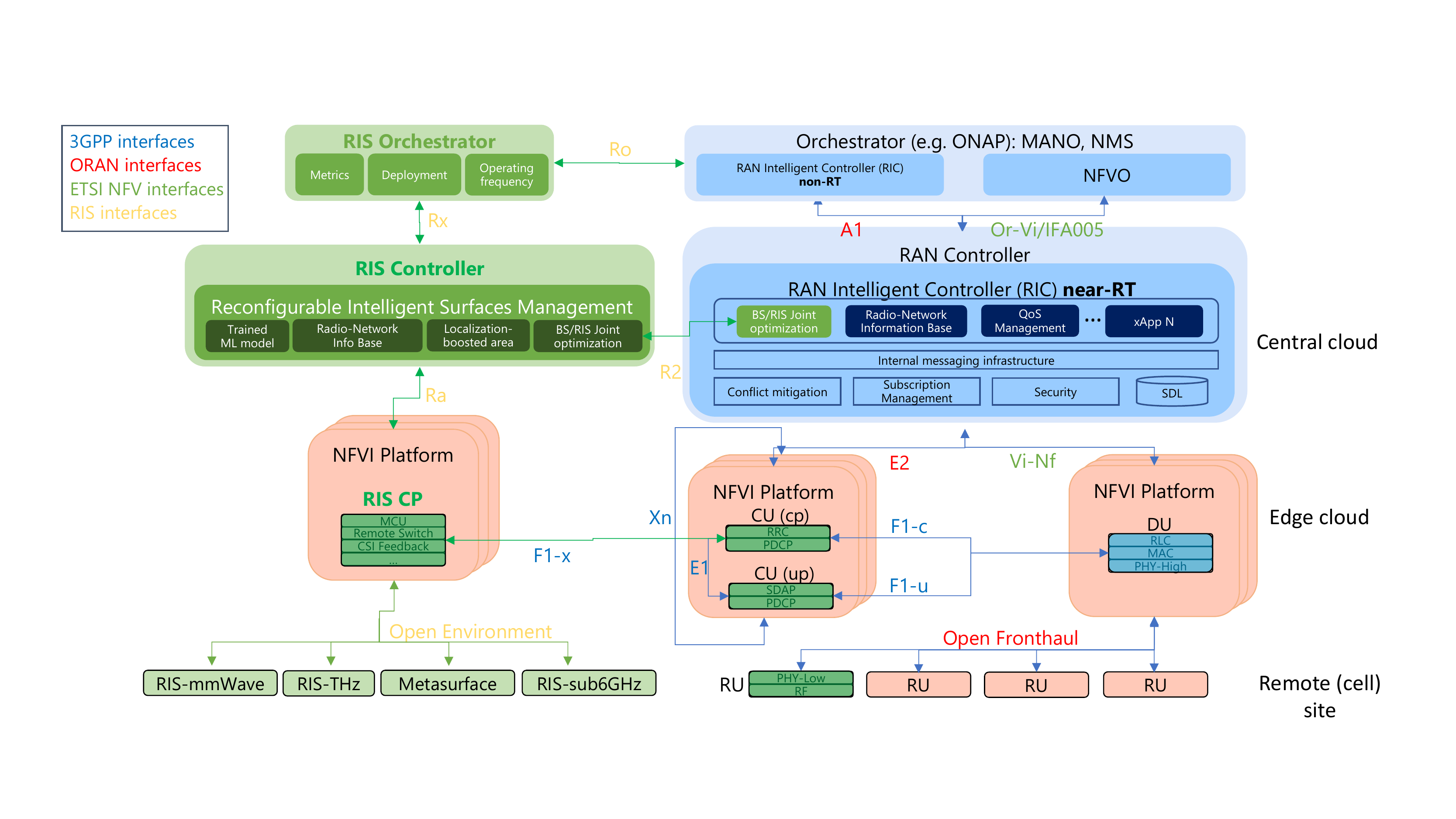}
    \caption{\small{The proposed network architecture integrating RISs with existing standardized interfaces, such as 3GPP, O-RAN, and ETSI.}}
    \label{fig:riseArch}
\end{figure*}

The right part of Fig.~\ref{fig:riseArch} illustrates the main functional blocks of existing network architectures, including Open RAN (O-RAN) and the RAN functional split according to 3GPP~\cite{rise6g_del25}. All 3GPP-compliant interfaces are colored in blue, whereas the O-RAN interfaces are red-colored. In the figure, we split the RAN elements, namely eNB or gNB according to the 3GPP jargon, specifically, into a Centralized Unit (CU), a Distributed Unit (DU), and a Remote Unit (RU). While DU functions can be placed on an edge cloud, they would require a very high capacity connection links to assist and support real-time operations between the Radio Link Control (RLC)/MAC layers and PHY. Additionally, the CU functions can be placed on edge clouds; in this case, they shall be split into Control Plane (CP) and User Plane (UP) functions. DUs are connected to RUs by means of the Open FrontHaul (O-FH) protocol. Following the O-RAN architecture perspective, the figure depicts the near-RealTime (near-RT) RAN Intelligent Controller (RIC) that automatically interacts with the CU functions through the indicated standardized E2 interface. This RAN controller can automatically integrate third-party applications, namely xApps, which control specific RAN functions. Also, the non-RealTime (nonRT) RIC can exchange messages with the near-RT RIC using the standardized A1 interface. According to our proposed RIS-enabled smart wireless environment architecture in Fig.~\ref{fig:riseArch}, the aforementioned conventional network architecture blocks can interact with new RIS-oriented network entities (i.e., the left part of the figure) via the novel interfaces represented by F1-x, R2, and Ro. Two representative use cases can be envisioned by means of the latter interfaces:
\begin{itemize}
    \item The RIS is directly connected to enB/gnB enabling BS transmit and RIS reflective BF to be jointly optimized, i.e., the network operator owns the RIS deployment. In this case, the CU control plane can trigger specific phase configuration at the RIS via its actuator within few milliseconds. It is noted that the RIS actuator will be responsible for actuating the logical commands received by the RIS controller, i.e., for translating them into physical configurations to be applied to the RIS panel. This direct connection of the RIS to the enB/gnB is particularly relevant when the RIS deployment is under the network operator control.
    \item The RIS is connected to the management of the eNB/gNB in a master/slave or peer-to-peer fashion. In this case, the RIS actuator and the RIS panel can only be configured by the RIS controller. However, the latter interacts with the near-RT RIC by means of a dedicated xApp that will have its counterpart in the RIS controller. This case includes self-contained and independent RIS deployments, which can be dynamically integrated into the network and used by operators upon request.
\end{itemize}
This open environment implementing RIS-enabled smart wireless networking is envisioned as a class of new protocols that can be optimized to include different types of RISs (e.g., reflective, transmissive, and simultaneous reflecting and sensing~\cite{Tsinghua_RIS_Tutorial}). 

\subsection{The RIS Bandwidth of Influence (BoI) Challenge} \label{sec:BoI_definition}
Consider a wall or any “dumb surface” in the signal propagation environment of a wireless network. The reflection properties of this wall depend on its material and the network's operating frequency $f$, as well as on the impinging angles of its incoming signals. In general, signal reflections from the wall follow Snell's law~\cite{RIS_Cap_Opt} (i.e., geometrical optics) and cannot be dynamically configured. In contrast, each unit cell of a reflective RIS has a reconfigurable reflection and/or transmissive capability, appearing in a limited range of frequencies of the impinging wave. This range of frequencies that an RIS can impact signal propagation (i.e., reflection/transmission) is defined as its BoI. In fact, an RIS is composed of a set of usually identical unit cells (called also as unit elements or meta-atoms), and its overall influence over wireless communications cannot exceed that of its single unit cell, which is limited in the frequency domain. Therefore, studying the BoI of a unit cell in a uniform RIS is sufficient to characterize this metric for the entire RIS. This will be the focus in the following. 
\begin{figure}[!t]
\centering
  \includegraphics[clip,scale=0.75]{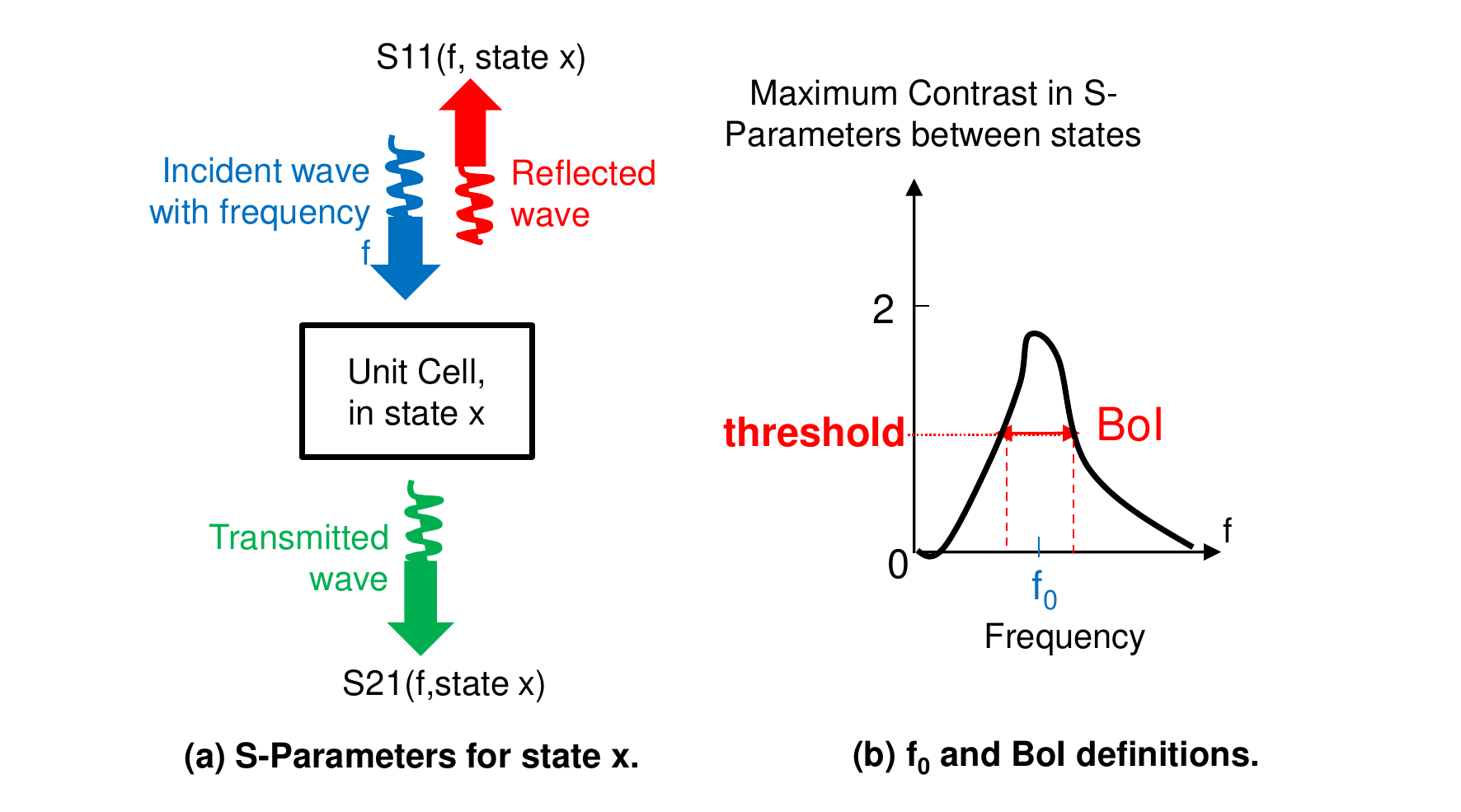}
  \caption{Schematic definition of the maximum contrast and the BoI of a unit cell of an RIS.}
  \label{fig:BoI_concept}
\end{figure}

As illustrated in Fig.~\ref{fig:BoI_concept}a, when the unit cell of an RIS, which is configured in a state $x$, gets illuminated by an impinging wave with unit power and frequency $f$, it reflects this wave with a complex reflection coefficient ${\rm S}11(f,x)$ and also transmits another wave with a complex transmission coefficient ${\rm S}21(f,x)$. For a given frequency $f$, these ${\rm S}$-parameters depend on the state $x$ of the unit cell. Let us define $C_R(f)$ and $C_T(f)$ as the maximum contrasts among all states of the unit cell in terms of reflection and transmission, respectively. Following these definitions, $C_R(f)$ and $C_T(f)$ can be mathematically expressed for any feasible unit-cell state $x$, and a feasible unit-cell state $x'\neq x$, as follows:
\begin{align*}
    C_R(f) &= \max_{\forall x,x\neq x'} \left\lvert {\rm S}11(f,x) - {\rm S}11(f,x') \right\rvert, \\
    C_T(f) &= \max_{\forall x,x\neq x'} \left\lvert {\rm S}21(f,x) - {\rm S}21(f,x') \right\rvert,
\end{align*}
where $\lvert \cdot \rvert$ denotes the amplitude of a complex number, and it holds that $0\leq \lvert C_R(f) \rvert \leq 2$ and $0\leq \lvert C_T (f)\rvert\leq2$ since the ${\rm S}$-parameters are bounded by $1$. It is noted that, when the maximum contrast values are both close to zero, the unit cell behaves like a ``dumb point'' (e.g., a point of a “dumb surface”), i.e., a non-reconfigurable one. In this case, its ${\rm S}$-parameters only depend on $f$ and the unit cell has no capability to change state, i.e., its state is fixed depending on its material and on its placement with respect to the impinging wave.
\begin{figure}[!t]
\centering
  \includegraphics[trim={5cm 11cm 2cm 5cm},clip,scale=0.70]{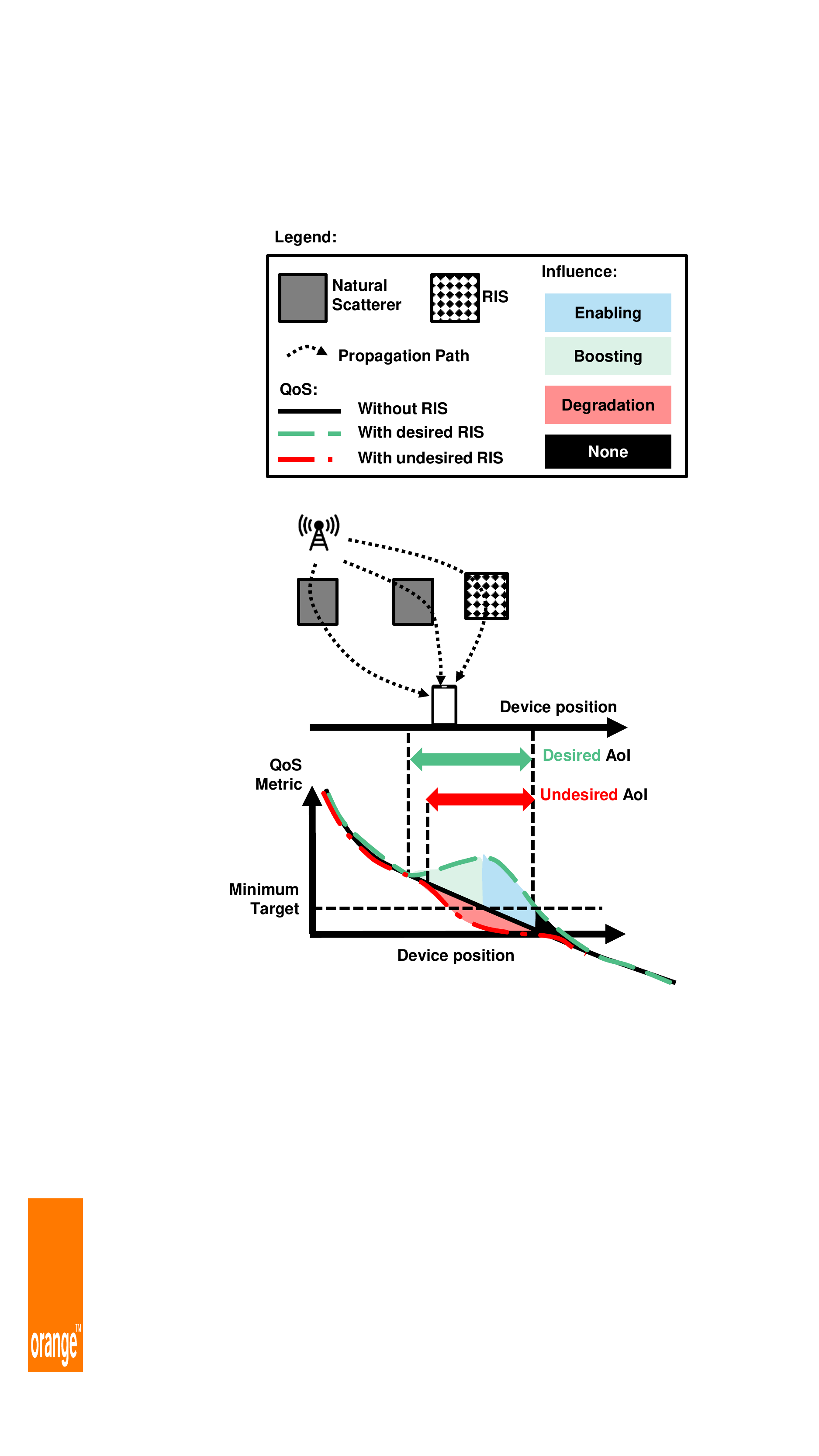}
  \caption{Illustrative description of the AoI of an RIS. The RIS may either improve or degrade the QoS in certain geographical areas, leading to desired or undesired AoI, respectively.}
  \label{fig:AoI_concept}
\end{figure}

Let us define a threshold $C_{\min}$ beyond which the maximum contrast $C_R (f)$, or $C_T (f)$, of a reflective RIS is considered as significant, and consequently, its unit cell can be considered as having a more significant dynamically adjustable influence on the wave that it re-radiates at frequency $f$ (due to an impinging wave at this frequency). As the contrast is lower bounded by $0$ and upper bounded by $2$, in the sequel, we chose $C_{\min}$ equal to $1$. The BoI of an RIS is defined as the range of frequencies for which the contrast is more significant for both its incoming (i.e., impinging) and outgoing (i.e., reflected or re-transmitted) wave. For instance, the unit cell of an efficient reflective RIS will have a more significant dynamically adjustable influence on the wave reflection inside its BoI, and a weaker influence on the reflection happening outside of it. Using the latter definitions and notations, the BoI of the unit cell of a reflective RIS can be mathematically expressed as follows (see Fig.~\ref{fig:BoI_concept}b for an illustrative interpretation):
\begin{equation*}
    \operatorname{BoI} \triangleq \{ f \vert C_R(f) \geq C_{\min} \},
\end{equation*}
where it should also hold that $C_R(f) \gg C_T(f)$. 
In Section~\ref{sec:BoI_Char} that follows, characterizations of contrasts and BoIs for various implementations of unit cells for one transmissive and three reflective RISs will be presented. For those investigations, we further introduce the RIS central frequency $f_0$. By representing the BoI as the frequency set $\left[f_1,f_2\right]$, $f_0$ is defined as follows: 
\begin{equation*}
    f_0 \triangleq \frac{f_1+f_2}{2}.
\end{equation*}
Note that alternative $f_0$ definitions could be used, e.g., $f_0 \triangleq \arg\max_{f \in \left[f_1,f_2\right]} C_R(f)$. 


\subsection{The RIS Area of Influence (AoI) Challenge} \label{sec:aroi_def}
As illustrated in Fig.~\ref{fig:AoI_concept}, the RIS can be configured to contribute an additional signal propagation path to the available paths between a BS and a UE. In this way, it can influence the existing coverage between those two nodes. Specifically, the RIS can either improve/extend the coverage, in terms of a Quality of Service (QoS) metric, or degrade it, depending on its phase profile setting. The geographical area where the coverage of a BS for a given service with a given target QoS can be modified via an RIS is defined as the AoI of the RIS. In particular, the area where the service coverage of the BS is improved, or extended, is the desired AoI, whereas the area where it is degraded is the undesired AoI. In mathematical terms, the AoI of an RIS of any type can be defined as follows:
\begin{equation*}
    \operatorname{AoI} \triangleq \{ \mathcal{S} \,\vert\, m(\mathcal{S}) \geq q_{\rm th} \},
\end{equation*}
where $m(\cdot)$ represents the desired performance metric evaluated over a geographical area $\mathcal{A}$ under investigation and $q_{\rm th}$ is the minimum service requirement. Note that it must hold: $\mathcal{\mathcal{S}\subseteq \mathcal{A}}$.

Neither the shape nor the location of the AoI of an RIS is intuitive, and both depend, among other factors, on the considered performance metric. In practice, a planning tool based on ray-tracing is necessary to visualize the AoI~\cite{Phan_Huy_2022}, and to contribute in choosing the best locations for deploying RISs~\cite{Albanese_2022_risaware}. In Section~\ref{sec:AoI_char} that follows, characterizations of AoIs for various types of RIS-boosted services will be discussed.

\begin{figure}[!t]
\centering
  \includegraphics[trim={0cm 3cm 11cm 0cm},clip,scale=0.70]{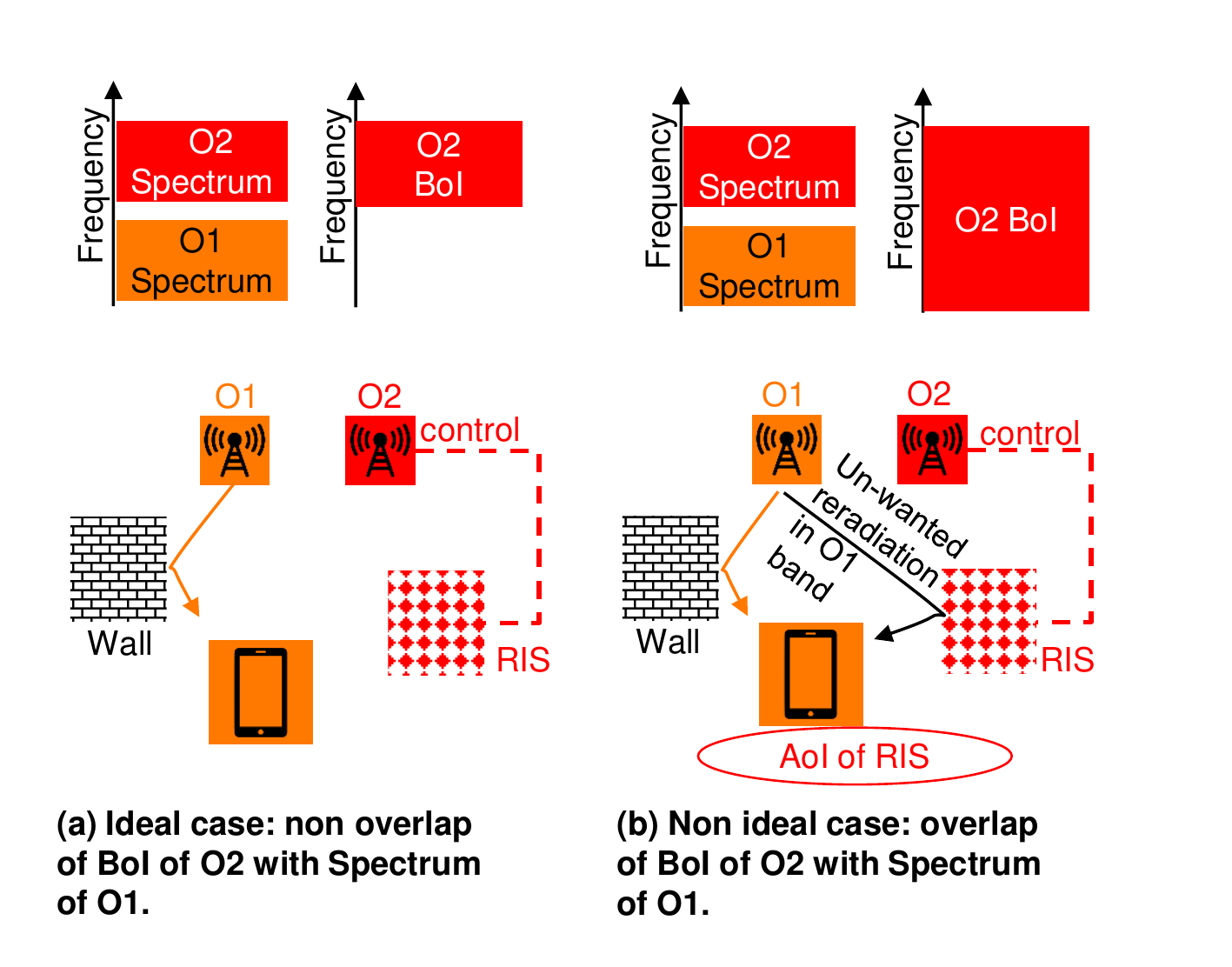}
  \caption{The operators' coexistence challenge when an RIS is deployed.}
  \label{fig:operator_coexist}
\end{figure}
\subsection{Operators Co-Existence with RISs} \label{sec:aroi_boi_impact}
Figure~\ref{fig:operator_coexist} illustrates the case where two network operators O1 and O2 possess different spectrum bands and each deploys a separate BS. It is assumed that operator O2 also deploys a reflective RIS. In Fig.~\ref{fig:operator_coexist}a, the BoI of the RIS is depicted as ideal, meaning that it perfectly matches the spectrum owned by O2, influencing solely its coverage, and not that of operator O1. However, in Fig.~\ref{fig:operator_coexist}b, the non-ideal case where the BoI of the RIS of O2 overlaps with the spectrum owned by O1 is considered. In this subfigure, we further assume that the RIS has an AoI that overlaps with the geographical coverage of O2. This implies that a UE served by operator O1 and located in the RIS's AoI will suffer from undesired reflections from the RIS. Those reflections are undesired, in the sense that, they fluctuate in a discontinuous and unpredictable manner for the service provided by O1. Such fluctuations may degrade the performance of closed-loop link adaptation mechanisms (such as adaptive modulation and coding, power control, and adaptive BF) as well as of fast scheduling schemes, degrading the overall link performance.
\begin{figure}[!t]
\centering
  \includegraphics[trim={0cm 1.5cm 11cm 0cm},clip,scale=0.70]{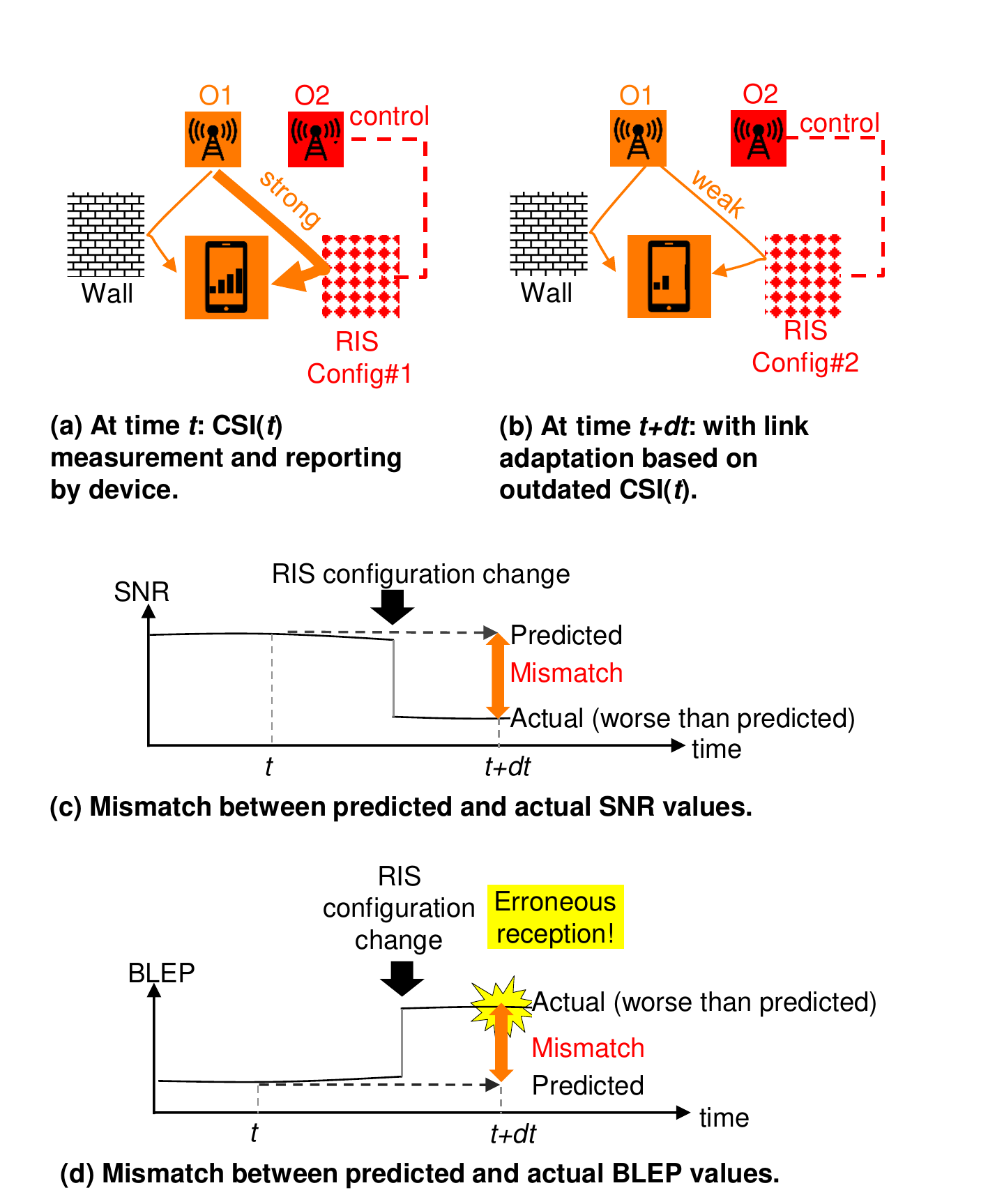}
  \caption{Channel state information mismatch due to time-varying undesired reflections from an RIS with a BoI and an AoI overlapping with the spectrum and coverage, respectively, of an unintended network operator.}
  \label{fig:CSI_mismatch}
\end{figure}

The non-ideal RIS deployment case in Fig.~\ref{fig:operator_coexist}b is investigated in more detail via the example illustrated in Fig.~\ref{fig:CSI_mismatch}. As shown in Fig.~\ref{fig:CSI_mismatch}a, a UE connected to the operator O1 acquires and reports the Channel State Information (CSI) at time instant $t$, which is represented as CSI($t$). At time instant $t+dt$, as depicted in Fig.~\ref{fig:CSI_mismatch}b, the UE receives data from O1 with a link adaptation based on CSI($t$). However, due to the existence of an RIS operated by O1, which changed from phase configuration $1$ to $2$ between instants $t$ to $t+dt$, respectively, the channel between the BS of O1 and the UE changed. Figure~\ref{fig:CSI_mismatch}c demonstrates that this RIS phase configuration change may suddenly degrade the received Signal-to-Noise Ratio (SNR) at the UE. This actually indicates a mismatch between the SNR prediction (based on the CSI($t$) value) used for link adaptation, and the actual SNR at time $t$. As depicted in Fig.~\ref{fig:CSI_mismatch}d, this translates into a large BLock Error Probability (BLEP) resulting in an erroneous packet reception.

To conclude, we showcased that an RIS deployed by a network operator, and having a BoI overlapping with the spectrum of another operator, can degrade the QoS of the latter. This fact witnesses the importance of the BoI characterization of RIS equipment in driving their efficient design and fabrication in order to ensure smooth co-existence among operators that deploy RISs.

\section{Bandwidth-of-Influence Experimental Characterization} \label{sec:BoI_Char}
In this section, we present novel experimental results for the BoI of the various fabricated RISs in the framework of the ongoing RISE-6G project. In particular, we characterize the BoI of the following RIS unit-cell implementations:  
\begin{itemize}
    \item Unit cell of an $1$-bit Transmissive RIS (T-RIS)~\cite{DiPalma_2016_1bit, DiPalma_2017_circularly,TRIS2023};
    \item Unit cell of a varactor-based Reflective RIS (R-RIS)~\cite{Fara_2022_prototype, Ratajczak_2009, Ratajczak_2010};
    \item Unit cell of a RF-switch-based R-RIS~\cite{Rossanese_2022_designing}; and
    \item Unit cell of a PIN-diode-based R-RIS.
\end{itemize}	
For each implementation, the characterization includes: \textit{i}) the maximum contrast (defined in Sec.~\ref{sec:BoI_definition}) as a function of the frequency $f$; \textit{ii}) the BoIs at the value $C_{\min}=1$, which serves as the contrast threshold; and \textit{iii}) the central operating frequency $f_0$ of the RIS structure. 

\subsection{BoI of an $1$-bit PIN-Diode-based Transmissive RIS} \label{sec:boi_T_RIS}
\begin{figure}[!t]
\centering
  \includegraphics[scale=0.50]{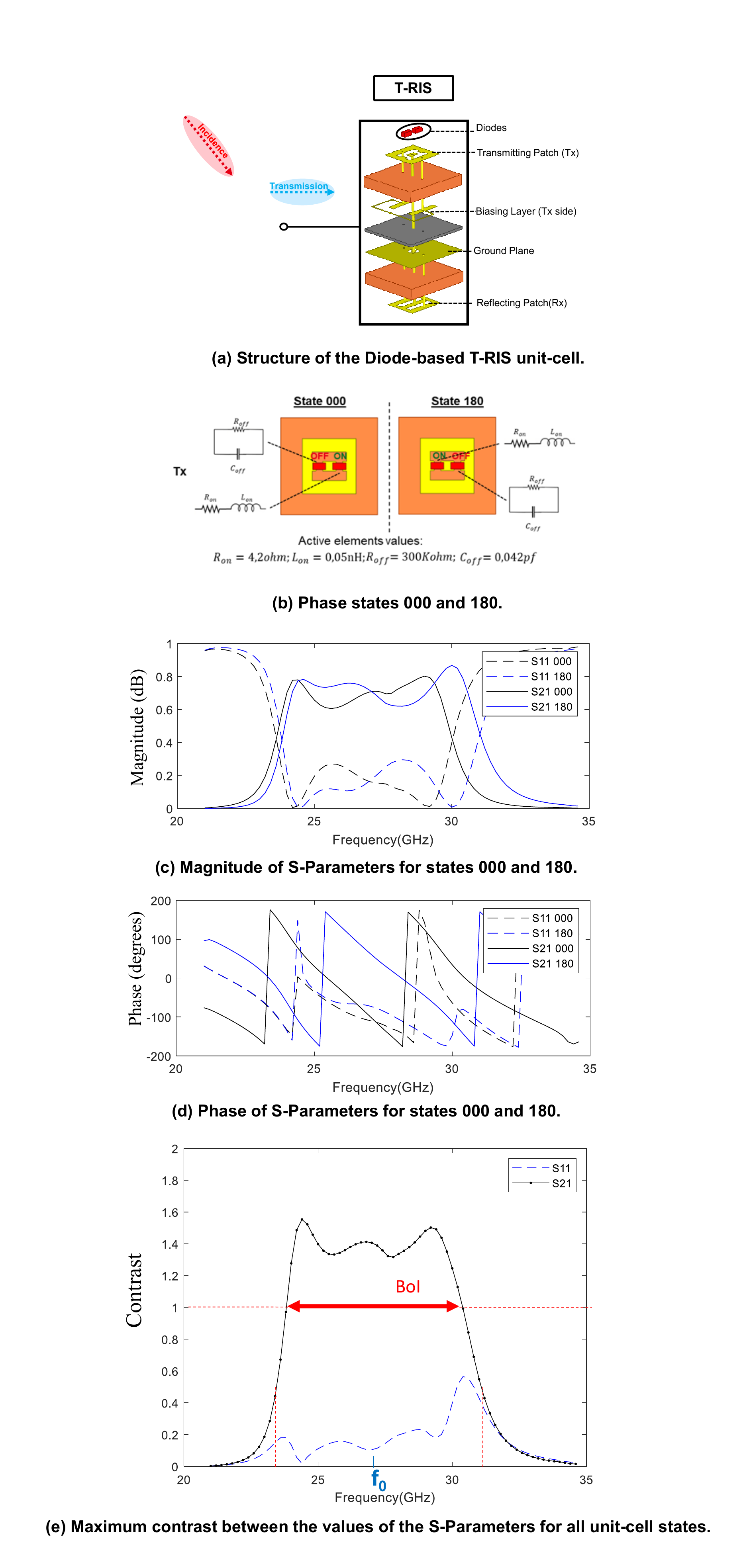}
  \caption{PIN-Diode-based T-RIS unit-cell characterization: (a) structure, (b) states $000$ and $180$ and the (c) magnitude and (d) phase of their ${\rm S}11$ and ${\rm S}21$ parameter, and (e) contrast of the ${\rm S}$-parameters for these states as a function of the frequency in GHz.}
  \label{fig:T_RIS}
\end{figure}
Figures~\ref{fig:T_RIS}a and~\ref{fig:T_RIS}b depict the structure of the unit cell of an $1$-bit T-RIS, which is an improved version of the unit cell firstly introduced in~\cite{DiPalma_2016_1bit, DiPalma_2017_circularly}. This improved unit cell has the following characteristics \cite{TRIS2023}: \textit{i}) it is designed to operate at frequencies lying approximately between $23$ GHz and $32$ GHz, and \textit{ii}) it is composed of the  four metal layers: receiving patches; transmitting patches; biasing lines (transmission sides); and ground plane.
This T-RIS is illuminated by a source from the receiving patch side. On the transmitting patch, the alternation of the diodes state produces a $180^{\rm o}$ phase difference. The unit cell ensures an $1$-bit phase quantization for transmission. Figures~\ref{fig:T_RIS}c--\ref{fig:T_RIS}e illustrate the magnitude, the phase, and the contrast (which is the difference between the two possible states of the unit cell) of the ${\rm S}$-parameters as functions of the frequency $f$ in GHz, as obtained by the Ansys HFSS full-wave EM simulator under periodic boundary conditions. For this $1$-bit unit cell, the contrast in terms of the reflection $C_R(f)$ and transmission $C_T(f)$ are computed as follows:
\begin{align*}
    C_R(f) &= \lvert {\rm S}11(f,180^o) - {\rm S}11(f,0^o) \rvert, \\
    C_T(f) &= \lvert {\rm S}21(f,180^o) - {\rm S}21(f,0^o) \rvert.
\end{align*}

In Fig.~\ref{fig:T_RIS}e, the contrasts of the ${\rm S}11$ and ${\rm S}21$ parameters are demonstrated versus different frequency values. As expected in this case (recall that this implementation is a T-RIS), $C_T(f)$ is much larger than $C_R(f)$, hence, we use $C_T(f)$ to determine the BoI (see Sec.~\ref{sec:BoI_definition} for the justification). It can be seen that the BoI at the maximum contrast threshold $C_{\min}=1$ is $[23.9, 30.6]$ GHz, resulting in a width of $6.7$ GHz with a central frequency $f_0$ equal to $27.3$ GHz.
   
\subsection{BoI of a Varactor-based Reflective RIS} \label{sec:boi_R_RIS}
\begin{figure}[!t]
\centering
  \includegraphics[trim={1cm 1cm 0.3cm 1cm},clip,scale=0.45]{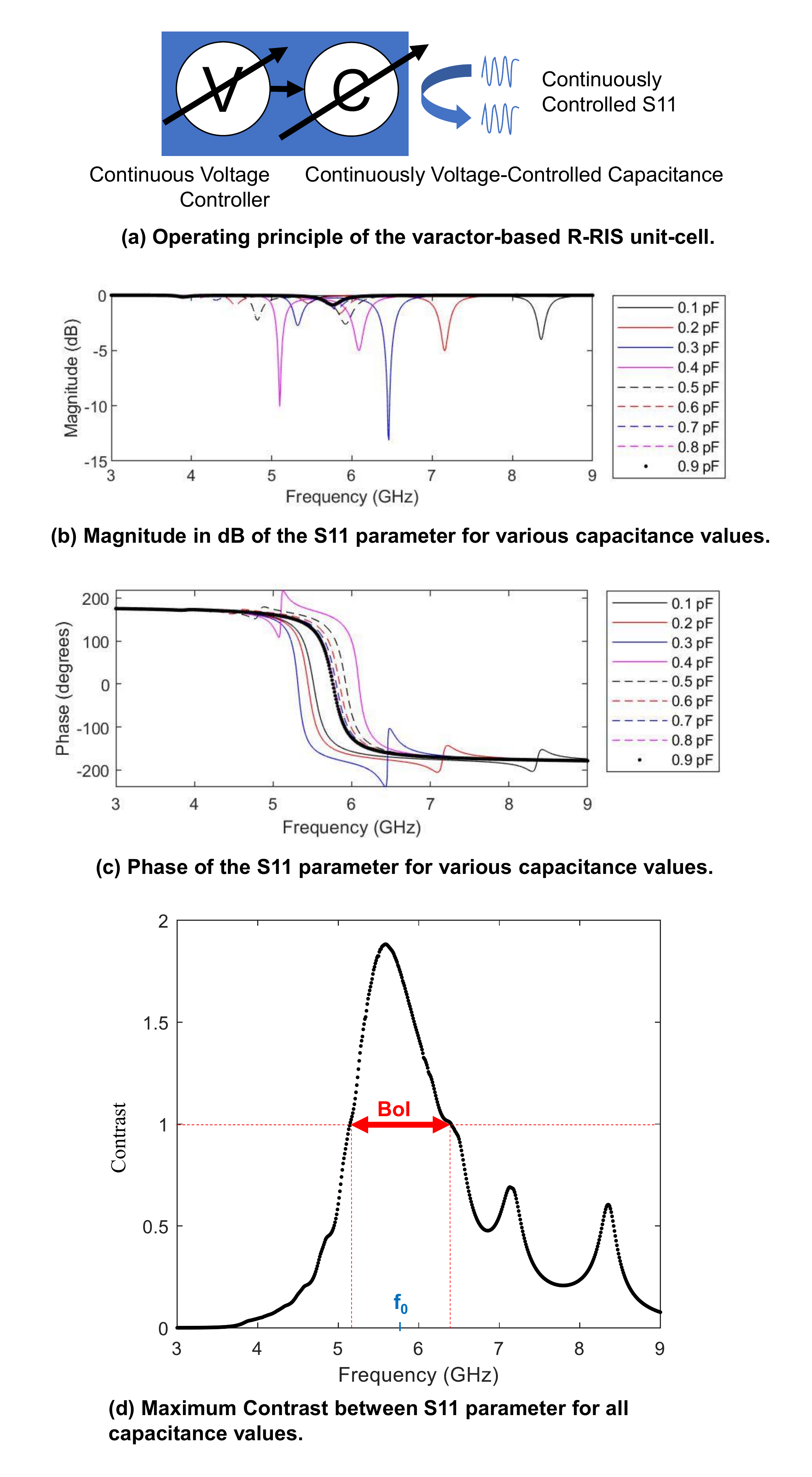}
  \caption{Varactor-based R-RIS unit-cell characterization: (a) operation principle, (b) magnitude and (c) phase of the ${\rm S}11$ parameter for various capacitances, and (d) contrast of the ${\rm S}11$ parameter for various capacitance values versus the frequency in GHz.}
  \label{fig:R_RIS}
\end{figure}
In this subsection, we characterize the unit cell of a varactor-based R-RIS with continuous phase shifting~\cite{Fara_2022_prototype, Ratajczak_2009, Ratajczak_2010}. As illustrated in Fig.~\ref{fig:R_RIS}a, this unit cell is loaded with a variable capacitance, whose values range from $0.1$ to $0.9$ pF, which is voltage-controlled continuously via diode-varactors~\cite{Mortenson1974variable}. Hence, the ${\rm S}11$ parameter accepts continuous values and its magnitude and phase, as functions of the frequency $f$ in GHz, are illustrated in Figs.~\ref{fig:R_RIS}b and~\ref{fig:R_RIS}c, respectively, considering $9$ different values of the capacitance between $0.1$ and $0.9$ pF with $0.1$ pF step. Figure~\ref{fig:R_RIS}d includes the maximum contrast $C_R(f)$, which was obtained for each possible pair of capacitance values. 
It can be observed that the BoI at the contrast threshold $C_{\min}=1$ is approximately $[5.1, 6.4]$ GHz, having a width of $1.3$ GHz. In this case, the central frequency $f_0$ is equal to $5.8$ GHz.

\subsection{BoI of an RF-Switch-based Reflective RIS} \label{sec:boi_RF}
\begin{figure}[!t]
\centering
  \includegraphics[trim={0cm 1cm 0cm 0cm},clip,scale=0.7]{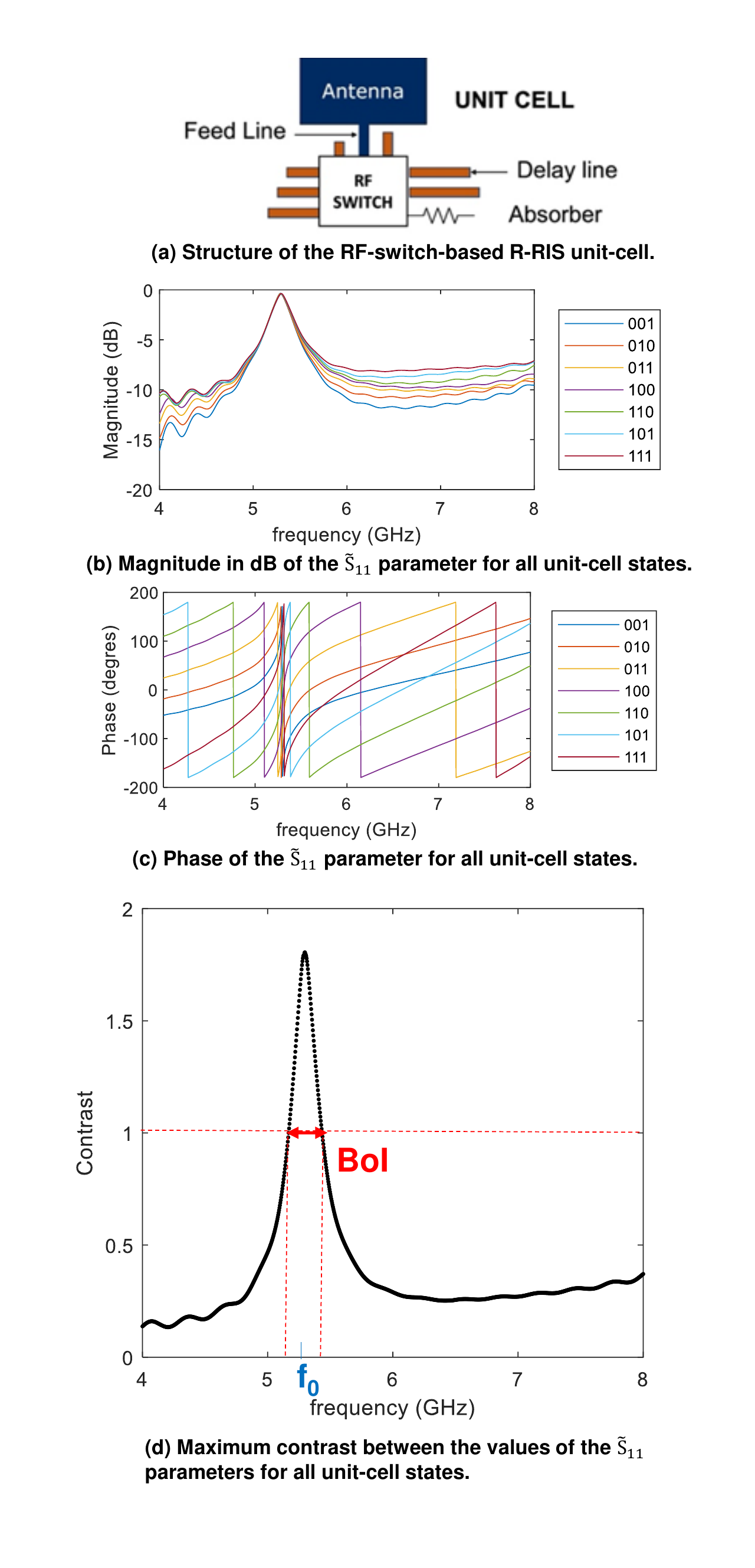}
  \caption{RF-switch-based R-RIS unit-cell characterization: (a) structure,  (b) magnitude and (c) phase of the ${\rm S}11$ parameter for the $7$ different configurations, and (d) contrast of the ${\rm S}11$ parameter for the considered different configurations versus the frequency in GHz.}
  \label{fig:RF_R_RIS}
\end{figure}
In this subsection, we evaluate the BoI of the unit cell of the RF-switch-based R-RIS in \cite{Rossanese_2022_designing}, which is depicted in Fig.~\ref{fig:RF_R_RIS}a. This unit cell was designed to operate at $5.3$ GHz and consists of a patch antenna connected to an RF switch, which is in turn connected to $8$ different outputs: $7$ of these outputs are attached to open-ended delay lines of different lengths, while $1$ of them is attached to a $50$ $\Omega$ matched absorber, which allows to effectively turn off the unit cell. The length of each delay line is designed to impose $7$ different phase shifts to the impinging signal at the unit cell; those phase shifts are equally spaced in the range from $0$ to $2\pi$. 

The magnitude of the ${\rm S}11$ parameter represents the portion of the energy of the impinging signal that is immediately reflected back at the interface with the patch antenna. Hence, in order to compute the BoI, we first define the following variable:
\begin{equation*}
    \delta = 1 - \lvert {\rm S}11 \rvert,
\end{equation*}
which represents the portion of the energy of the impinging signal that enters the patch antenna, gets re-directed to one of the delay lines of the unit cell, and finally gets reflected back in the air. We, thus, define the unit cell ``effective ${\rm S}11$'' as follows:
\begin{equation*}
    \tilde{{\rm S}}11 = \delta e^{\jmath \angle {\rm S}11},
\end{equation*}
where $\angle {\rm S}11$ represents the argument of the complex number ${\rm S}11$. Finally, for every frequency value $f$, we compute the maximum contrast in the effective ${\rm S}11$ as:
\begin{equation*}
    C_R(f) = \max_{\forall c_i,c_i \neq c_j} \lvert \tilde{{\rm S}}11(f,c_i) - \tilde{{\rm S}}11(f,c_j) \rvert,
\end{equation*}
where $c_i$ and $c_j$ represent any two distinct RIS unit-cell configurations.

The magnitude and phase of the $\tilde{\rm S}11$ parameter, as obtained from full-wave simulations, are shown in Figs.~\ref{fig:RF_R_RIS}b and~\ref{fig:RF_R_RIS}c, respectively. It is depicted that the magnitude of $\tilde{\rm S}11$ exhibits a strong positive peak at the working frequency of $5.3$ GHz, meaning that most of the incoming signal's energy is reflected back by the unit cell's patch antenna. Moreover, at this frequency, the phase of $\tilde{\rm S}11$ attains $7$ equally spaced values in the range from $0$ to $2\pi$. Note that, in this case, the BoI never goes to zero since, even outside the operating frequency range, a (small) portion of the impinging signal always enters the unit cell and gets reflected through a different delay line, depending on the chosen configuration. It is noted that this effect can be mitigated by tweaking the design parameters of the unit cell (e.g., its dimensions).
Finally, Fig.~\ref{fig:RF_R_RIS}d demonstrates the BoI at the contrast threshold $C_{\min}=1$ which is $[5.17, 5.44]$ GHz, yielding a width of $270$ MHz. The central frequency $f_0$ for this R-RIS fabricated structure is $5.3$ GHz.

\subsection{BoI of an $1$-bit PIN-Diode-based Reflective RIS} \label{sec:boi_pin_R_RIS}
\begin{figure}[!t]
\centering
  \includegraphics[trim={1cm 0cm 0cm 0cm},clip,scale=0.80]{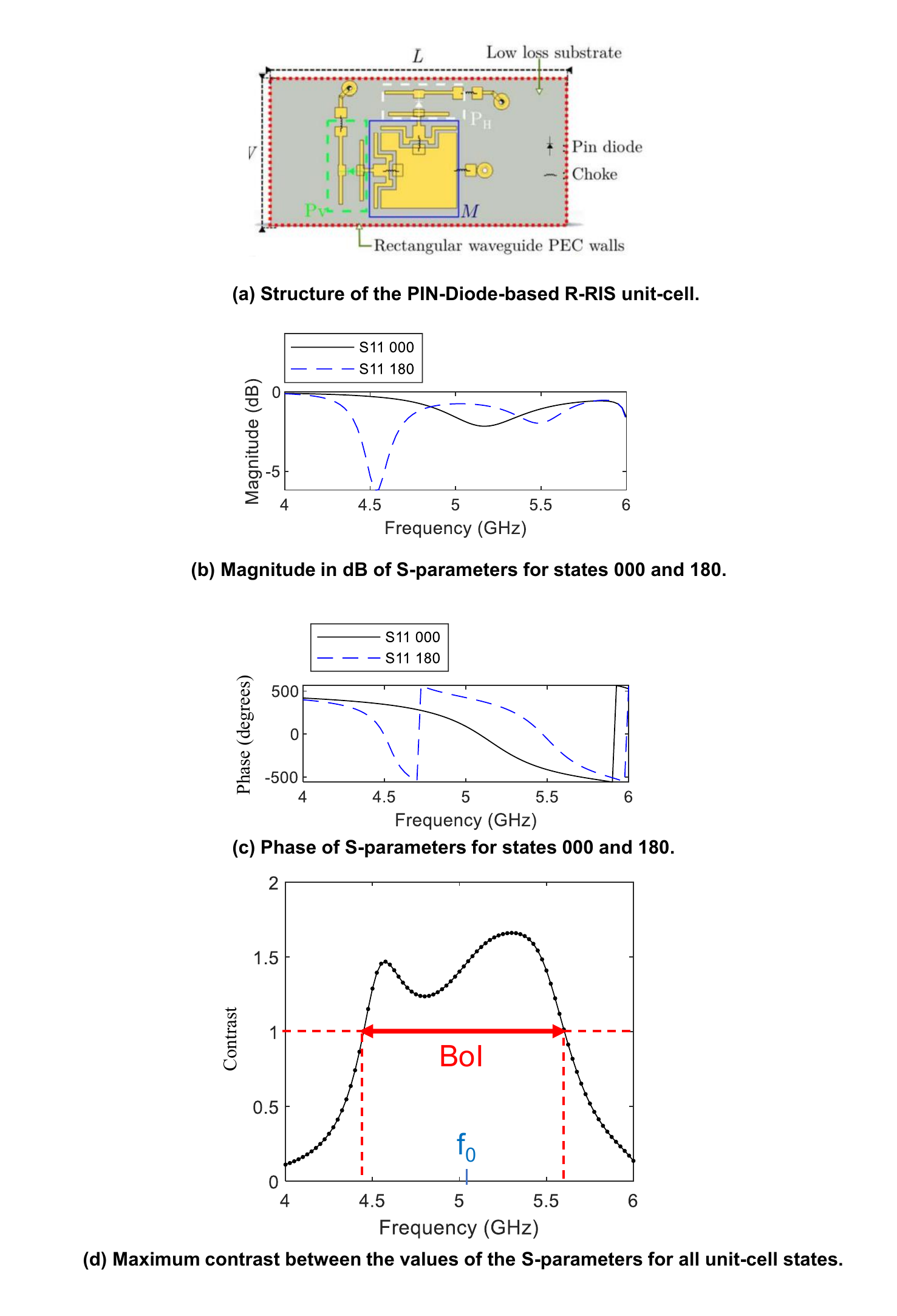}
  \caption{PIN-diode-based R-RIS unit-cell characterization: (a) structure, (b) magnitude and (c) phase of the ${\rm S}11$ parameter for states $000$ and $180$, and (d) contrast of the ${\rm S}11$ parameter for these states as a function of the frequency in GHz.}
  \label{fig:Pin_Diode}
\end{figure}
Hereinafter, we evaluate the BoI of the unit cell of an $1$-bit PIN-diode-based R-RIS operating in the sub-6GHz frequency band. The schematic view of this cell as well as the active elements values corresponding to its two possible states are illustrated in Fig.~\ref{fig:Pin_Diode}a. The designed dual linearly-polarized unit cell operates at the central frequency $5.2$ GHz. It is composed of one metal layer consisting of two orthogonal folded dipoles and a matching network circuit printed on a thin $0.1$ mm FR4 substrate. This substrate, characterized by $\epsilon_r=4.3$ and $\tan(\delta)=0.01$, is placed upon a thick $3$mm LERD spacer (with $\epsilon_r=1.04$ and $\tan(\delta)=0.0001$) and is backed by a metallic $1$ mm ground plane realized with an aluminium reflector. At the top layer, two radiating elements are present, which are placed normal to each other to cope two orthogonal polarizations. Two PIN diodes are connected to the inner printed matching network and are biased in opposite states; one diode is switched ON, while the other one is in OFF state.

By operating in this binary mode, it is possible to obtain a reflection coefficient phases of $0^{\rm o}$ or $180^{\rm o}$ degrees, depending on the polarization state of the diodes. This $1$-bit unit cell has been simulated for its two-phase states using the commercial software Ansys HFSS with periodic boundary conditions on the cell's lateral faces and Floquet port. As shown in Fig.~\ref{fig:Pin_Diode}c, the unit cell exhibits a phase difference of almost $180^{\rm o}$ over the whole considered frequency band for both states. The unit cell's magnitude is depicted in Fig~\ref{fig:Pin_Diode}b for both the $0^{\rm o}$ and $180^{\rm o}$ states. The contrast in terms of the ${\rm S}11$ parameter is plotted as a function of the frequency $f$ in GHz in Fig.~\ref{fig:Pin_Diode}d. As it can be observed, the BoI at the contrast threshold $C_{\min}=1$ is approximately $[4.4, 5.6]$ GHz with a width of $1.2$ GHz. In this case, the central frequency $f_0$ is defined as the mid-frequency of the BoI and is equal to $5.05$ GHz.
    
\subsection{BoI Comparison among Different RIS Implementations} \label{sec:boi_compare}
\begin{figure}[!t]
\centering
  \includegraphics[scale=0.85]{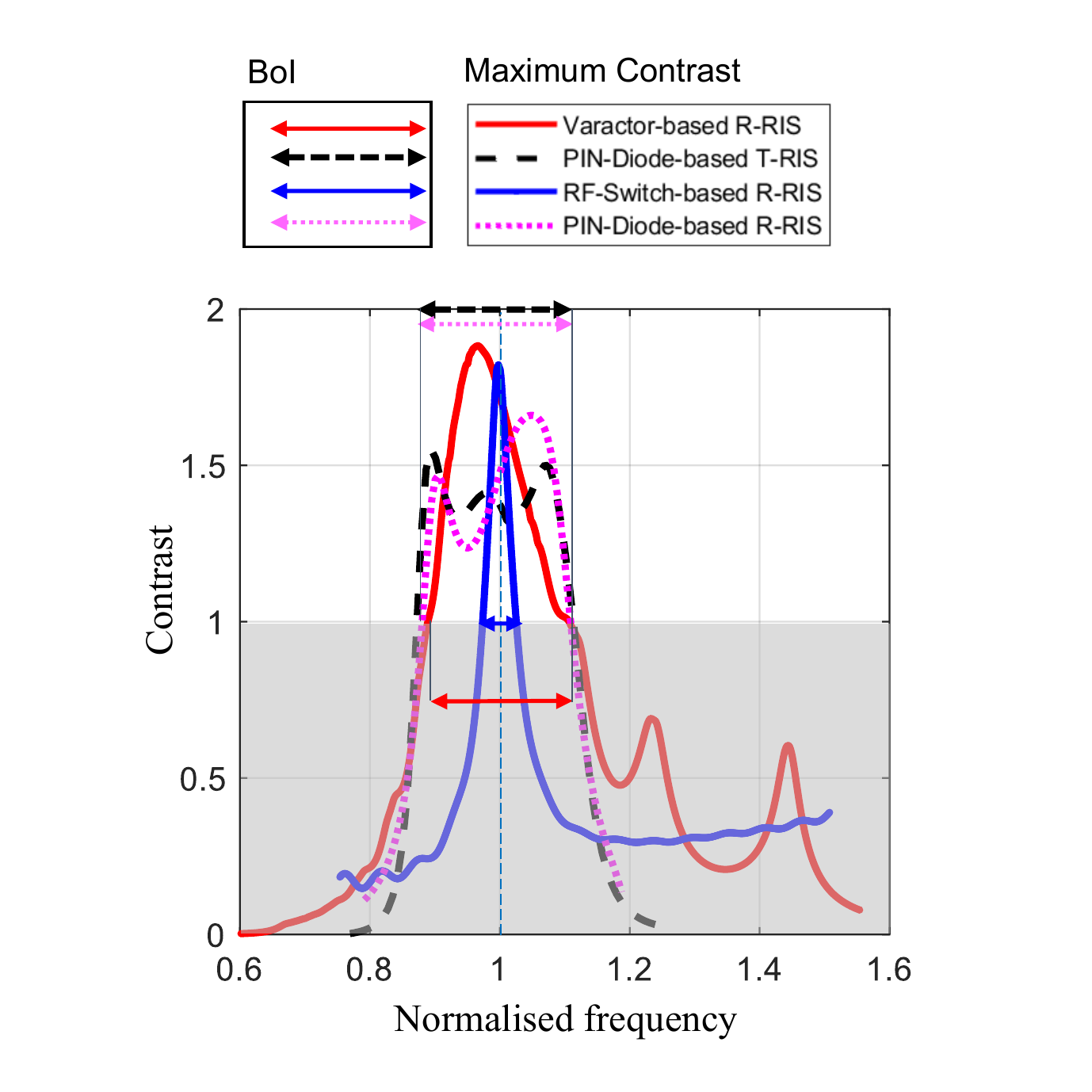}
  \caption{Normalized BoI for the four different fabricated RIS unit cells.}
  \label{fig:Compare_BoI}
\end{figure}
In Fig.~\ref{fig:Compare_BoI}, we compare the BoIs at the contrast threshold $C_{\min}=1$ for the four aforedescribed fabricated RIS unit cells. In particular, the maximum contrasts as a function of the normalized frequency, i.e., the frequency divided by the central operating frequency $f_0$, are illustrated. As observed, the varactor-based R-RIS and the PIN-diode-based RISs exhibit the largest BoI, while the RF-switched-based R-RIS results in the smallest BoI. 

It is expected that the resonating frequency of an RIS unit cell depends on its physical and/or electrical size, as well as its state. Hence, its behavior, in terms of contrast versus frequency, depends on the implementation of the reconfiguration mechanism. It can be observed, for the unit cell of the PIN-diode-based R-RIS, that the maximum contrast is quite flat within its BoI, while dropping abruptly outside of it. This is due to the fact that, when the unit cell is on the ON state, it has a slightly larger physical size, hence, it strongly resonates in the lower part of the BoI, but still significantly resonates in its upper part. However, when the unit cell is on the OFF state, it has a slightly smaller physical size, but it again strongly resonates in both the upper and lower parts of the BoI. This implies that the BoI's lower and higher frequencies are close enough, yielding a nearly flat maximum contrast. A similar contrast behavior appears to the unit cell of the PIN-diode-based T-RIS. In this case, however, the lower part of the BoI is mainly due to the patch's size (which slightly increases), whereas BoI's upper part is mainly due to the slot size in the patch (which slightly decreases). In this RIS unit-cell design, the reconfiguration only changes the orientation of the current circulating on the patch.  

Regarding the maximum contrast behavior for the unit cells of the RF-switch-based and varactor-based R-RISs, it is reminded that the former's states result from changes on the length of the cell's feeding line, whereas the latter offsets, in the frequency domain in a continuous manner, its peak resonant frequency linked to its electrical size. To this end, in the former case, the BoI depends on the resonant frequency of the patch antenna's physical size (recall that this antenna resonates in a narrow band), whereas, in the latter case, the contrast behavior over frequency is mainly limited by the cell's physical size.   

In conclusion, different hardware designs and fabrication approaches of RIS unit cells yield different BoIs, which should be carefully considered when deploying RISs in wireless networks, as discussed in Sec.~\ref{sec:aroi_boi_impact}. Evidently, an RIS with a large BoI is not always the best option. In fact, the RIS hardware design, in terms of its BoI, needs to be optimized according to the intended deployment scenario and use case. 


\section{Area-of-Influence Experimental Characterization} \label{sec:AoI_char}
In this section, we investigate the desired AoI of a reflective RIS for several use cases referring to different QoS requirements. In particular:
\begin{itemize}
    \item Boosting a UE's EE using an RIS; in this use case, the self exposure of the UE to its own emissions is also reduced. This implies that this use case can be also considered as an S-EMFEU boosting use case.
    \item RIS-enabled localization boosting.
    \item RIS-enabled SSE boosting.
\end{itemize}

\subsection{Example of an EE-/S-EMFEU-Optimized AoI} \label{sec:Aroi_EE_SEMFEU}
\begin{figure}[!t]
\centering
  \includegraphics[trim={2cm 9cm 4cm 8cm},clip,scale=1]{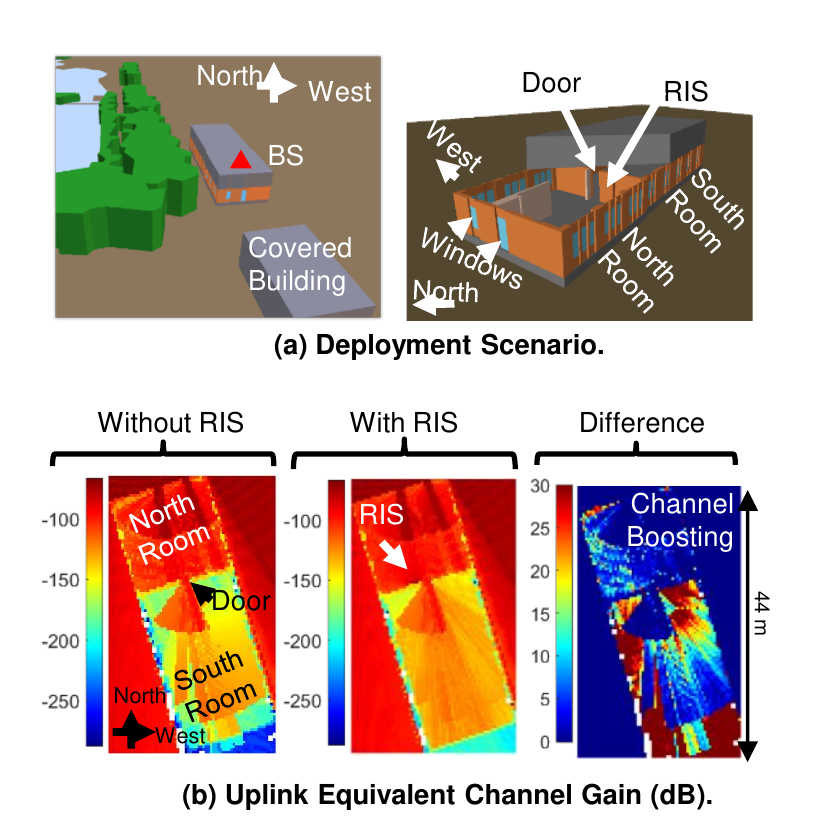}
  \caption{The simulated scenario for different UE positions, aiming at the EE/S-EMFEU AoI characterization. The coverage inside and outside the building is analysed to unveil the influence of the RIS deployment. It is showcased that, deploying a reflective RIS in the south room, leads to improvements in terms of achieving higher channel gains within this room.}
  \label{fig:Simul_scenario}
\end{figure}

We first focus on the use case of EE/EMFEU boosting and connectivity enabling. More precisely, we consider a deployment scenario where a UE transmits information to an outdoor massive Multiple-Input Multiple-Output (MIMO) BS (i.e., in the uplink) from inside or outside a building, in the absence or presence of a reflective RIS. This deployment scenario has been simulated using a ray tracing tool; the detailed simulation assumptions and the methodology followed are available in~\cite{Phan_Huy_2022}. In Fig.~\ref{fig:Simul_scenario}a, we analyse the coverage inside and outside a building with two rooms: a ``north room'' with two north windows and a ``south room'' with only one north door. The massive MIMO BS was located outside north. To characterize the influence of the RIS deployment, we have compared two scenarios: one without the RIS and another one with the RIS placed in the north room upon the south wall. 

Let us first visualize how the RIS improves the propagation between the UE and the BS. Figure~\ref{fig:Simul_scenario}b illustrates three heatmaps of the equivalent channel gain between the UE (for all positions of the UE upon the map) and the BS, taking into account the BS's reception BF: 
\begin{itemize}
    \item The heatmap of the channel gain in the absence of the RIS;
    \item The heatmap of the channel gain in the presence of the RIS with joint optimization of the BS receive BF and the RIS reflective BF; and
    \item The heatmap of the channel gain improvement due to the introduction of an RIS. 
\end{itemize}
As depicted in the figure, without the RIS, the channel is particularly strong close to the two north windows of the north room, and very poor in the south room, except to the area close to the door between the two rooms. Note that outside the building the channel is extremely strong. The RIS has been intentionally positioned on the south room, on a location which benefits from a strong channel, thanks to one of the north windows. The RIS mainly improves the gain in the south room, by creating propagation paths between the south room and the BS.

\begin{figure}[!t]
\centering
  \includegraphics[trim={2cm 8cm 4cm 7cm},clip,scale=1.1]{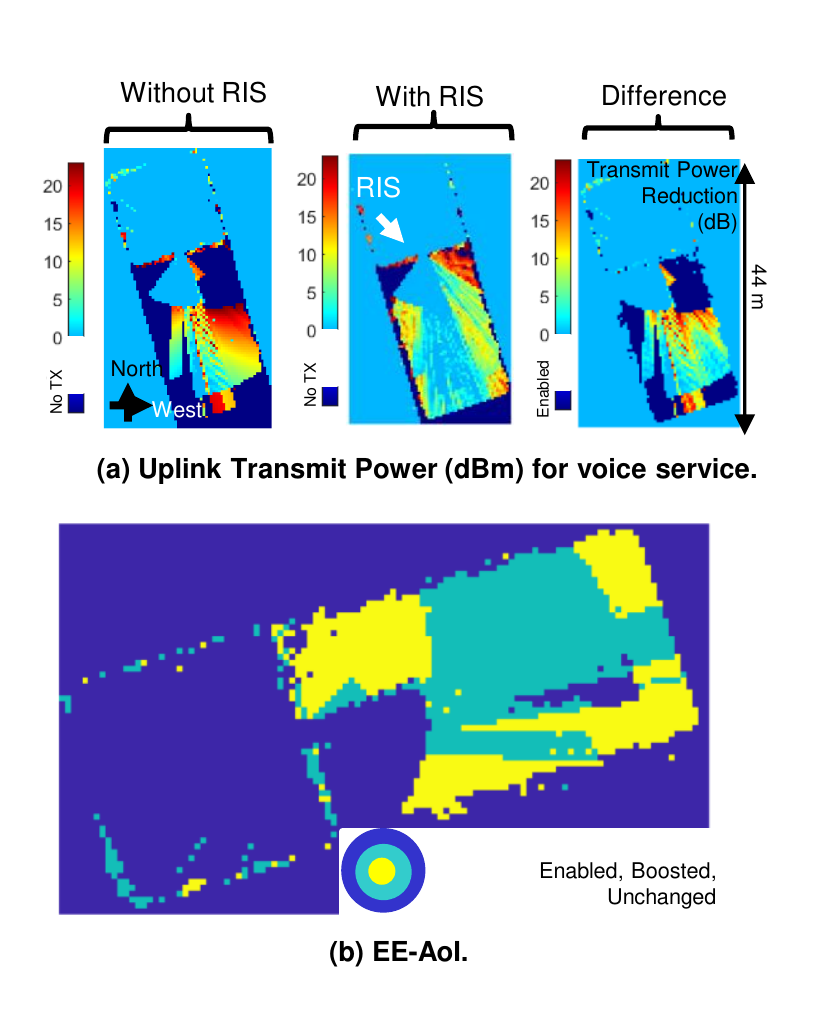}
  \caption{EE-/S-EMFEU-optimized AoI, where three types of influences are illustrated: the ``unchanged'' (blue region), the ``boosted'' (green region), and the ``enabled'' areas (yellow region) showcasing the spatial distribution for the EE/S-EMFEU metric.}
  \label{fig:EE_AoI}
\end{figure}

We next visualize how the improvements in channel gain, illustrated in Fig.~\ref{fig:Simul_scenario}b, translate into improvements for a given performance metric, such as the EE from the UE perspective. We now consider the use case where the UE wishes to transmit voice data to the BS. Figure~\ref{fig:EE_AoI}a illustrates three heatmaps of the UE transmit power adapted to the propagation channel to provide voice service, with a target receive SNR at the BS: \textit{i}) the UE transmit power in absence of the RIS; \textit{ii}) the UE transmits in the presence of the RIS whose reflective BF is jointly optimized with the BS receive BF; and \textit{iii)} the UE transmit power reduction (i.e., EE improvement) due to the introduction of an RIS. It is demonstrated that, when the propagation environment is poor, the voice QoS cannot be attained even with maximum power. This implies that the UE does not transmit any data since it is out of coverage. 

Figure~\ref{fig:EE_AoI}b illustrates three types of areas: \textit{i}) the ``unchanged area'' where no change due to the RIS is observed (the UE remains out of coverage or keeps reaching the maximum data rate enabled by the standard with the minimum transmit power enabled by the device); \textit{ii}) the ``boosted area'' where the UE transmit power is reduced and the EE is boosted; and \textit{iii}) the ``enabled area'' where the UE gets voice coverage thanks to the RIS. The EE-optimized AoI is the combination of the ``boosted-area'' and the ``enabled area''. It is noted that the EMFE of the UE to its own emissions is also reduced when its transmit power is reduced. Therefore, the S-EMFEU-boosted area is the same as the EE-boosted area, and the same holds for the S-EMFEU and EE AoIs.

\subsection{Example of an Uplink SE-Optimized AoI} \label{sec:Aroi_SE_uplink}
\begin{figure}[!t]
\centering
  \includegraphics[trim={2cm 7.5cm 4cm 7.5cm},clip,scale=1.1]{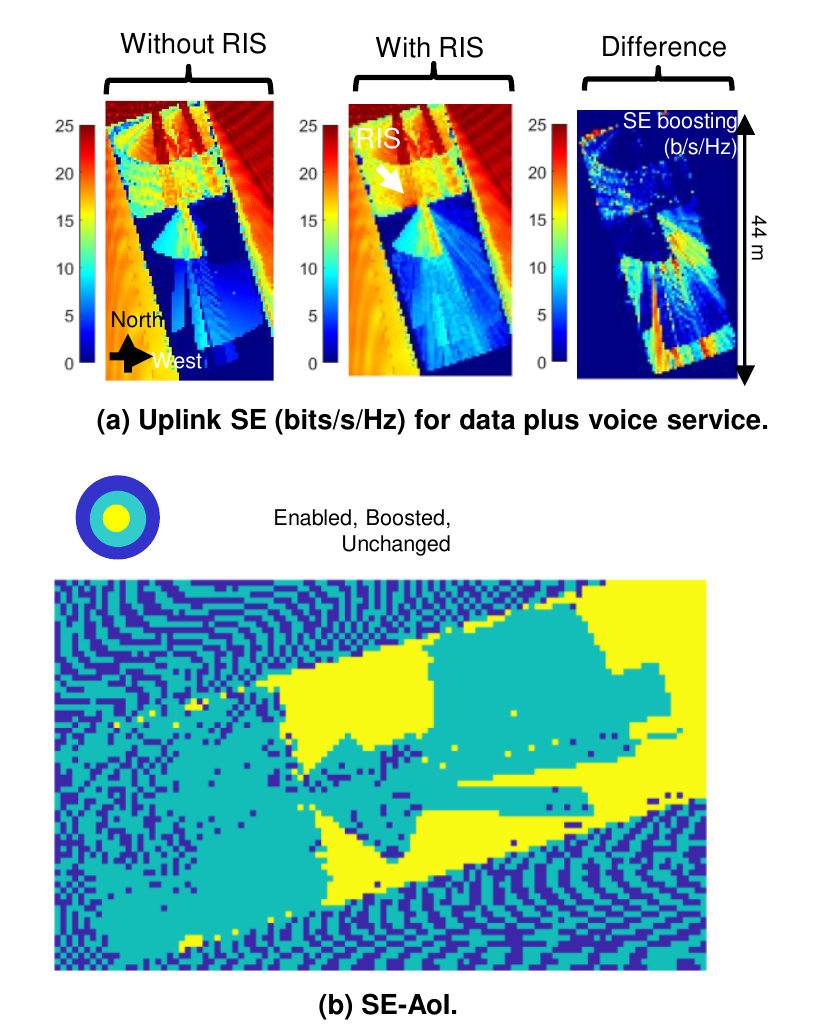}
  \caption{The RIS-enabled AoI optimization for the uplink SE metric for the case where the UE is transmitting using its maximum available power.}
  \label{fig:SE_AoI}
\end{figure}

In this section, we visualize how the improvements in channel gain illustrated in Fig.~\ref{fig:Simul_scenario} do translate into improvements for the SE performance metric, expressed in bits/sec/Hz. We consider a use case where the UE transmits voice service plus best effort data traffic using its maximum transmit power. Figure~\ref{fig:SE_AoI}a illustrates the uplink Spectral Efficiency (SE) when the UE transmits with maximum power and adapts the data rate (thanks to adaptive modulation and coding) to the propagation channel, with the aim to maximize the best effort service data rate (in addition to the voice service). Three heatmaps are plotted in Fig.~\ref{fig:SE_AoI}b: \textit{i}) the SE without the reflective RIS; \textit{ii}) the SE in the presence of the RIS with jointly optimized BS receive BF and RIS reflective BF; and \textit{iii}) the SE improvement due to the introduction of RIS. It is shown that, when the propagation conditions are weak to offer the minimum performance (i.e., the voice service alone), the UE does not transmit at all being out of coverage.

Similar to Section~\ref{sec:Aroi_EE_SEMFEU}, three types of areas are depicted in Fig.~\ref{fig:SE_AoI}b: \textit{i}) the ``unchanged area'' where no change due to the RIS deployment and optimization is observed (the UE remains out of coverage or keeps reaching the maximum data rate enabled by the standard with the minimum transmit power enabled by the device); \textit{ii}) the “boosted area” where the SE is boosted; and \textit{iii}) the ``enabled area'' where the UE gets voice (and potentially data) coverage thanks to the RIS. The SE-optimized AoI is the combination of the ``boosted-area'' and the ``enabled area''.

\subsection{Example of a Localization-Optimized AoI} \label{sec:Aroi_localiz}
\begin{figure}[!t]
\centering
  \includegraphics[trim={2cm 9cm 2cm 2.5cm},clip,scale=1.0]{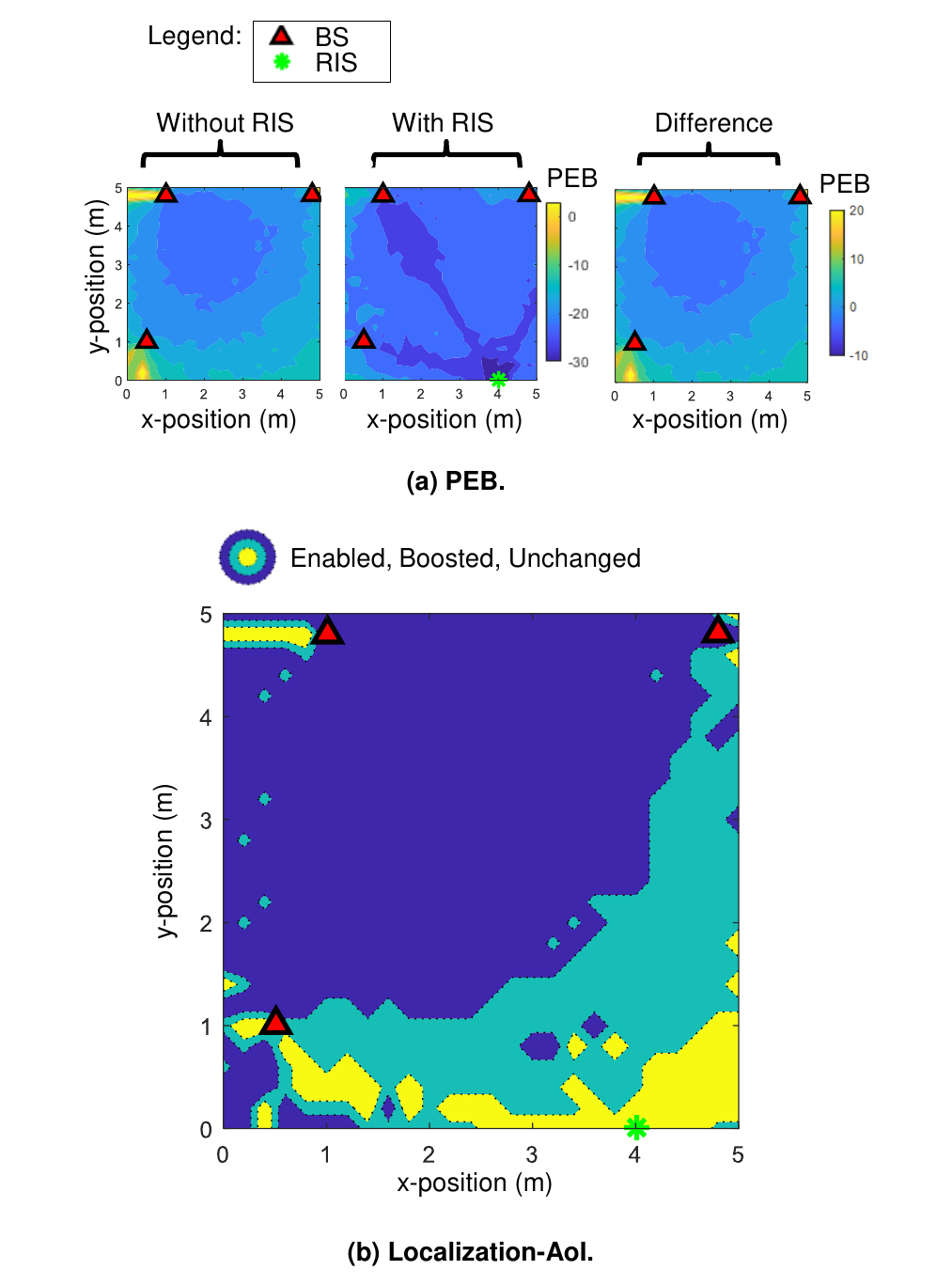}
  \caption{Localization AoI, showcasing the impact -in terms of the PEB (in dB)- of a reflective RIS deployed close to the lower right corner of the investigated scene (green star) to assist a conventional downlink positioning system relying on $3$ BSs (red triangles). Interestingly, the RIS enables localization and, even significantly boosts, accuracy in sub-regions close to the RIS (resp. yellow and green areas), without degrading performance elsewhere (blue area).}
  \label{fig:Localization_AoI}
\end{figure}

In Fig.~\ref{fig:Localization_AoI}, we deploy three single-antenna BSs (red triangles) at the points $[0.5, 1]^\top$, $[4.8, 4.8]^\top$, and $[1, 4.8]^\top$ in meters and a single-antenna UE to perform downlink multicarrier $2$D localization depending on three direct paths. A Uniform Linear Array (ULA) RIS in reflective mode (green star) deployed at the point $[4, 0]^\top$ in meters is installed to investigate its effect on the positioning performance. In our evaluation scenario, the RIS is assumed to be activated and controlled to produce one reflected path in the geometric Near-Field (NF) regime, in addition to the LoS direct path. This is achieved by exploiting the wave front curvature~\cite{Rahal_2021_all} and occurs exclusively when the BS closest to the RIS (located at $[0.5,1]^\top$) is transmitting. The remaining two BSs are considered to contribute only with direct paths to the system. Moreover, the end-to-end RIS precoder is randomized, but yet constrained to a predefined lookup table of realistic hardware RIS element-wise complex reflection coefficient measurements (the set $\mathcal{K}2$ in \cite{ Rahal_2022_arbitrary} with $2$-bit phase control and corresponding phases $0,\pi/2,\pi,-\pi/2$; another option could be the prototype in \cite{Fara_2022_prototype}), and a directional beam pattern is thus approximated.
However, the RIS performance can be further improved by adopting a location-optimal phase design as in~\cite{Rahal_2022_constrained} instead of randomly drawing from the predefined lookup table. As for the simulation parameters, the BSs were considered to transmit $40$ pilots in a wireless environment with free-space pathloss. The operating frequency was $28$ GHz and the bandwidth was assumed sufficiently large with $256$ subcarriers and a $240$ KHz bandwidth per subcarrier. The simulation area was $5\times5$ square meters and the UE's position was assumed known at each point where the Fisher Information Matrix (FIM) was computed, after which the Cramer-Rao lower bound (CRLB) on each position estimation as derived in both scenarios (with and without the RIS).

Figure~\ref{fig:Localization_AoI}a provides heatmaps of the resulting Position Error Bound (PEB) without and with RIS, as well as the difference between these two cases. Figure~\ref{fig:Localization_AoI}b then shows, in three colors, the different effects of deploying a reflective RIS for localization purposes in the aforementioned scenario. First, the localization is said to be “infeasible” in the conventional (three BSs only) system, if the expected accuracy is arbitrarily worse than $0.1$ m (according to some a priori application requirements). Accordingly, the first area, in yellow, is called “enabled” by the RIS, if the target accuracy is then met (in the same region). We can observe this effect, more particularly, in the area closest to the RIS, a.k.a in the AoI, where positioning is not just enabled ,but actually enhanced up to a considerable level. We also see this effect behind the BSs, where a poor GDoP would deprive the conventional system from properly inferring the UE location. The second effect that we notice, in blue, is labeled as ``unchanged.'' This is where the three BSs already achieve rather a good localization performance, while the RIS does not have a noticeable improvement (less than $2$ dB). This effect is mostly visible in the inner region between the three BSs. Lastly, we notice  a ``boosting'' effect in the green region, which is also known in the literature as a ``RIS-boosted localization'' regime.  Accordingly, while the conventional system relying uniquely on the three BSs can achieve acceptable performance, the RIS further contributes in improving localization to a significant extent, which leads to PEB level improvement by $3$ dB or more.

\subsection{Example of an SSE-Optimized AoI} \label{sec:Aroi_SSE}
In this section, we consider a scenario including a legitimate communication pair, comprising a BS and an RX, as well as an eavesdropper Eve, with all nodes equipped with multiple antennas and placed in the same geographical area.
\begin{figure}[!t]
\centering
  \includegraphics[scale=0.60]{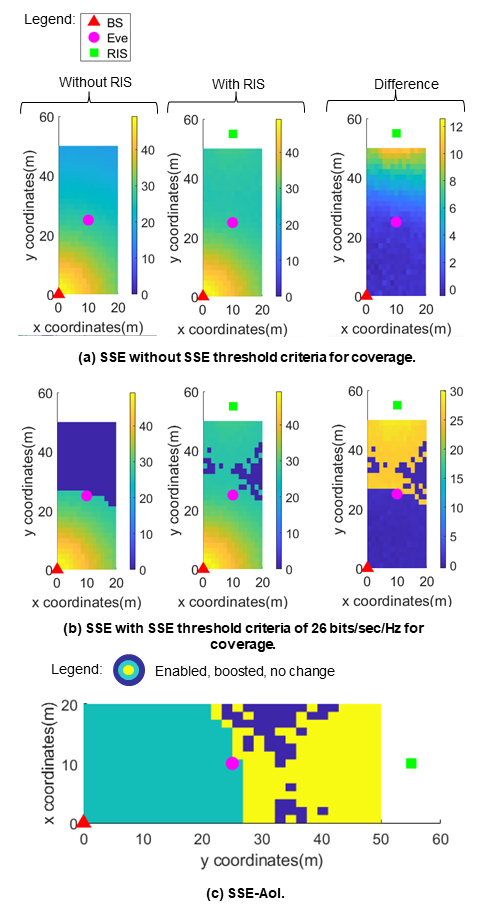}
  \caption{The RIS AoI for the SSE performance metric considering the presence of an eavesdropper (Eve) by: (a) not applying and SSE threshold, (b) imposing the SSE threshold of $26$ bits/sec/Hz, and (c) illustrating the resulting SSE-optimized AoI.}
  \label{fig:SSE_AoI}
\end{figure}

In Fig.~\ref{fig:SSE_AoI}a, we plot three different heatmaps of the SSE (which is defined as the data rate attained at the target UE minus the data rate that is intercepted by Eve \cite[eq. Sec.~II]{PLS_Kostas_all}): \textit{i}) when there is no RIS deployed; \textit{ii}) when there exists an RIS optimized for SSE maximization; and \textit{iii}) the difference in SSE performance due to the introduction of the RIS. More precisely, each heatmap is obtained assuming that the BS’s, Eve’s and RIS’s positions are fixed. In particular, the $3$D Cartesian coordinates of the BS, Eve and RIS were chosen as $[0,0,10]$, $[10,25,1.5]$, and $[10,55,5]$ in meters, respectively. Then, for each position of the legitimate RX on the map, the secrecy rate maximization problem was solved based on the design in \cite{PLS_Kostas_all,PLS2022_counteracting}, where the linear precoding and the artificial noise covariance matrix at the BS, the linear combining matrix at RX, and the RIS passive beamforming vector, as well as the number of the transmitted streams are optimized either for perfect or imperfect knowledge of Eve's CSI. In this figure's results, the BS was assumed equipped with $8$ antennas, Eve and RX share had both $4$ antenna elements, and the RIS consisted of $400$ unit cells. 

Figure~\ref{fig:SSE_AoI}b is derived from Fig.~\ref{fig:SSE_AoI}a by applying a minimum SSE threshold criterion, whose value was set equal to $26$ bits/sec/Hz. In locations where the SSE of Fig.~\ref{fig:SSE_AoI}a is below this predefined threshold, the network is incapable to serve the UE (as it cannot guarantee the minimum target SSE), thus, the UE is out of coverage the SSE was set to zero in Fig.~\ref{fig:SSE_AoI}b. Again, there exist three heatmaps of SSE in this figure: \textit{i}) without RIS; \textit{ii}) with RIS; and \textit{iii}) the difference due to the introduction of RIS. Figure~\ref{fig:SSE_AoI}c is derived from the right sub-figure of Fig.~\ref{fig:SSE_AoI}b and enables us to visualize three areas: an ``enabled area'' where the RX is served by the BS (with the guaranteed minimum SSE) thanks to the RIS deployment and proposed optimization; a ``boosted area'' where the SSE of RX is boosted thanks to RIS; and an ``unchanged area'' where RX remains out of the BS's coverage.

In the case where the RIS is not deployed, we observe in Fig.~\ref{fig:SSE_AoI}a that, close to the BS (i.e., at the bottom left corner) the secrecy rate is larger, reaching its maximum values throughout the considered grid of RX positions. This behavior is expected, since at these positions, Eve’s channel matrix is more degraded than RX’s. However, when RX is located away from the BS and specifically above Eve in the grid (i.e., when $y_{\rm RX}>y_{\rm Eve}$), SSE decreases considerably, reaching its minimum value, which is equal to $16.69$ bits/sec/Hz. On the other hand, when the RIS is deployed at the described position, it can be observed from the middle subfigure in Fig.~\ref{fig:SSE_AoI}a that the upper sub-area (i.e., for $y_{\rm RX}>y_{\rm Eve}$) gets boosted in terms of SSE performance. Particularly, the SSE increases approximately at least $2$ bits/sec/Hz close to Eve’s position and at least $10$ bits/sec/Hz to RIS’s position. The performance fluctuation of these two cases is better shown in the right subfigure Fig.~\ref{fig:SSE_AoI}a, where the difference between the latter two cases has been evaluated. In addition, it is observed that the maximum SSE difference, which is equal to $12.58$ bits/sec/Hz, is attained close to the position where the RIS is located, while for the sub-area close to the BS, the performance remains unchanged. In Fig.~\ref{fig:SSE_AoI}c, the SSE-optimized AoI is demonstrated, including areas where the targeted metric is either enabled or boosted, or remains unchanged.

\section{Conclusions} \label{sec:concl}
In this paper, we presented various deployment scenarios for RIS-enabled smart wireless environments, as they have been recently identified by the ongoing EU H2020 RISE-6G project. In particular, the proposed scenarios have been categorized according to the expected achievable performance, and their goal was to lead in substantial performance improvement compared to conventional network deployment architectures. However, similar to any promising new technology, the increased potential of RISs is accompanied with certain challenges. Two novel challenges with RISs are the BoI and AoI, which were introduced and mathematically defined in this article, as well as characterized via preliminary simulations results and measurements. For the former challenge, it was showcased that it depends on the considered RIS unit-cell technology, requiring careful consideration especially when RISs are intended for multi-operator deployment cases. On the other hand, it was demonstrated that the AoI depends on the RIS hardware design and placement, and it can be dynamically optimized to meet certain performance objectives.

\bibliographystyle{IEEEtran}
\bibliography{references}

\end{document}